\UseRawInputEncoding 
\documentclass[twocolumn]{aastex7}
\usepackage{CJK}
\usepackage{amsmath}
\usepackage{graphicx}
\usepackage{breqn}

\newcommand{\teff}{$T_{\rm eff}$}

\newcommand{\Dnu}{$\Delta \nu$}
\newcommand{\logg}{\mbox{$\log g$}}

\usepackage{comment}

\begin{document}

\title{Data-Driven Ages from Asteroseismology for FGK Dwarfs}



\author[0000-0002-1910-5641]{Christopher Lam}
\affiliation{IPAC, California Institute of Technology, Pasadena, CA 91125, USA}
\affiliation{Department of Astronomy, University of Florida, Gainesville, FL 32611, USA}
\email{lamchris@ipac.caltech.edu}

\author[0000-0003-0012-9093]{Aida Behmard}
\affiliation{Center for Computational Astrophysics, Flatiron Institute, Simons Foundation, 162 Fifth Ave, New York, NY 10010, USA}
\email{abehmard@flatironinstitute.org}

\author[0000-0002-3247-5081]{Sarah Ballard}
\affiliation{Department of Astronomy, University of Florida, Gainesville, FL 32611, USA}
\email{sarahballard@ufl.edu}

\author[0000-0003-0174-0564]{Andrew R.\ Casey}
\affiliation{Center for Computational Astrophysics, Flatiron Institute, Simons Foundation, 162 Fifth Ave, New York, NY 10010, USA}
\email{acasey@flatironinstitute.org}
\affil{School of Physics and Astronomy, Monash University, Wellington Road, Clayton VIC 3800 Australia}

\author[0000-0002-4818-7885]{Jamie N.\ Tayar}
\affiliation{Department of Astronomy, University of Florida, Gainesville, FL 32611, USA}
\email{jtayar@ufl.edu}

\author[0000-0003-1012-3031]{Jared~A.~Goldberg}
\affil{Department of Physics and Astronomy, Michigan State University, East Lansing, MI 48824, USA}
\affiliation{Center for Computational Astrophysics, Flatiron Institute, Simons Foundation, 162 Fifth Ave, New York, NY 10010, USA}
\email{goldstar@msu.edu}

\begin{abstract}
Age is the one of the most difficult stellar properties to constrain. However, upcoming surveys such as PLATO are about to revolutionize our ability to measure FGK dwarf ages with asteroseismology. Here, we use The Cannon to construct a data-driven model for inferring asteroseismic-quality age measurements from large spectroscopic survey observations and demonstrate that forthcoming asteroseismology datasets from future surveys can be leveraged, together with our model, to obtain stellar age samples larger than can be obtained by those surveys alone. We train The Cannon on 323 FGK dwarf stars with asteroseismic age labels from NASA \textit{Kepler} photometry and high-resolution spectra from the Apache Point Observatory Galactic Evolution Experiment (APOGEE). We apply our model to FGK dwarfs from the \textit{Kepler} survey that exhibit solar-like oscillations and reproduce their reported asteroseismic-quality ages with an RMS scatter of 1.0 Gyr. We then apply our model to the sample of 3810 \textit{Kepler} FGK dwarfs with APOGEE spectra to infer data-driven asteroseismic ages. We repeat this for the larger sample of 48,828 FGK dwarfs with APOGEE spectra. Our work demonstrates that data-driven methods can robustly reproduce asteroseismic ages for FGK dwarfs. These methods will be invaluable tools for extending asteroseimology measurements from upcoming surveys (e.g., PLATO) to much larger stellar samples.
\end{abstract}

\section{Introduction} \label{sec:intro}

Stellar ages are vital for unraveling the formation history of our Galaxy, as well as stars and their planetary systems therein. The fidelity to which we can interrogate these processes is limited by stellar ages. Large stellar samples with age measurements are often derived from isochrones, which are less reliable for stars that evolve more slowly, e.g., FGK dwarfs. Isochrone-derived ages for these stars have uncertainties commensurate with the ages themselves due to the isochrones' tightly packed tracks in this region of the Hertzsprung-Russell (HR) diagram \citep{sahlholdt_benchmark_2019}. Other techniques have advantages and shortcomings that make them more competitive for age-dating some stars over others. Gyrochronology, which infers stellar ages based on their expected rotation spin-down, is relatively powerful at measuring the ages of younger stars, but less effective for stars without spots and stars older than the spin-down stall \citep{skumanich_time_1972, angus_toward_2019, curtis_when_2020, lu_this_2024, bouma_ages_2024}. Lithium abundance has been shown to correlate with stellar age as Li is transported and burned within stars' hot convective zones over its first few hundred Myr, but older than the lithium depletion boundary at $\sim$500 Myr, there is no longer enough Li with which to mark stellar age \citep{galindo-guil_lithium_2022, wood_tess_2023}. One of the most effective age-dating methods is by tagging stars based on membership in co-natal populations (e.g., clusters, associations). However, this method is limited to stars that can be traced back to an open cluster, which becomes increasingly difficult on Gyr timescales as stars interact with the Galactic potential and other stars, thereby losing their birth kinematic information \citep{dias_updated_2021}. Thanks to \textit{Gaia} DR3 \citep{vallenari_gaia_2023}, it is now also possible to infer dynamical ages on a population level using stellar kinematics, with dynamically hotter stars being generally older \citep{sagear_zoomies_2024, schmidt_resonant_2024}. However, meaningful dynamical ages remain difficult to obtain for individual stars. 

Asteroseismology is another stellar age measurement technique. It is capable of measuring individual stellar ages to within 10-15\% precision \citep{lebreton_stellar_2008, soderblom_ages_2010, chaplin_asteroseismology_2013}, and is not limited to specific age ranges, unlike other methods, e.g., gyrochronology \citep{angus_toward_2019, lu_this_2024, bouma_ages_2024}. Asteroseismic ages are measured from internal stellar oscillations, the frequencies and modes of which map to the interior sound speed of the star and allow us to deduce many of its properties, including interior temperature and pressure, mass, and age \citep{aerts_asteroseismology_2010}. Specifically, ages can be deduced from asteroseismic measurements in two ways. Ideally, precisely identifying the modes associated with the measured frequencies allows the small and large frequency separations ($\delta\nu$ and \Dnu, respectively) to be resolved. $\delta\nu$ is particularly sensitive to age because it probes the hydrogen mass fraction in the core as stars burn their hydrogen fuel and decrease in density, while \Dnu\ maps more strongly to mass, although both separations encode information about the other parameter (for more background, we refer the reader to Figure 1.11 and Section 1 of \citealt{aerts_asteroseismology_2010}, as well as \citealt{valle_relevance_2020}). 

However, $\delta\nu$ is more difficult to obtain than \Dnu, and while \textit{Kepler} was able to obtain $\delta\nu$ for some main sequence stars (e.g., \citealt{aguirre_standing_2017}), most asteroseismic ages of main sequence \textit{Kepler} stars have come from another method of using asteroseismology to deduce ages, which is to combine stellar evolution models with 
two key global parameters: $\nu_{max}$ (the frequency of maximum amplitude of the pressure mode) and \Dnu\ (the large frequency separation). These parameters, derived from time-series photometry, are related via scaling relations \citep{ulrich_determination_1986, bellinger_stellar_2019} to stellar radius, mass, and density, with which stellar evolution models can produce an age estimate. 

Asteroseismology is most effective for stars with high signal-to-noise, high-precision light curves sampled with high cadence over a long baseline. 
With its high-precision photometry and high cadence over nearly four years, the NASA \textit{Kepler} mission has enabled asteroseismology for tens of thousands of stars, most of which are evolved giants \citep{gilliland_kepler_2010, pinsonneault_apokasc_2014, pinsonneault_apokasc-3_2025}. \textit{Kepler} has also enabled asteroseismology for the brightest main-sequence and sub-giant stars, but because the internal oscillation frequencies of small stars are significantly more difficult to detect than those of giant stars, the sample of \textit{Kepler} FGK dwarfs with asteroseismic ages is small. \citealt{serenelli_first_2017} produced ages for 415 dwarf and sub-giant FGK stars, while \citealt{aguirre_standing_2017} produced ages for 66 main-sequence FGK stars. Since most exoplanet hosts stars are FGK dwarfs, expanding the sample of stars with asteroseismic measurements to the dwarf regime will be invaluable for investigating planet formation and evolution.

In 2027, ESA will launch the PLAnetary Transits and Oscillations of stars (PLATO) space telescope. PLATO will have 24 `normal' cameras with a cadence more than twice as fast as that of \textit{Kepler}'s short cadence (25 seconds, compared to 58.5 seconds), in addition to two `fast' cameras with an even faster cadence of 2.5 seconds \citep{rauer_plato_2014, rauer_plato_2024}. Together with a selection function that favors stars brighter than those observed by \textit{Kepler}, this will enable PLATO to measure precise solar-like oscillations for 245,000 FGK dwarf stars \citep{montalto_all-sky_2021, rauer_plato_2024, boettner_exoplanets_2024,goupil_predicted_2024}. 

We can potentially use the forthcoming PLATO data releases as training sets to infer ages for even more stars using their spectra. Specifically, the intersection of PLATO targets with APOGEE spectra will enable the training of a model on stellar spectra and asteroseismic labels that can then infer labels given other spectra. Previous studies have used data-driven methods to extend stellar age measurements to larger samples using spectroscopic information \citep{anders_spectroscopic_2023, leung_variational_2023, ciuca_chasing_2024, almannaei_towards_2026, tayar_evaluating_2026}. Of these, studies using The Cannon include \citealt{ness_spectroscopic_2016}, which used a data-driven model from The Cannon to infer stellar masses and ages for \textit{Kepler} red giant stars with APOGEE spectra, and \citealt{lu_similarities_2022}, which used The Cannon and APOGEE spectra to relate the $\alpha$-abundances of giant stars throughout the Milky Way to their ages. 
In this study, we train The Cannon on APOGEE spectra and asteroseismic labels of stars from the \cite{serenelli_first_2017} sample, and then demonstrate that the labels -- including age -- inferred by the model from the APOGEE spectra of other stars are both accurate and precise. 

This manuscript is organized as follows. In Section \ref{sec:the-cannon}, we provide an overview of The Cannon. In Section \ref{sec:training_sample}, we describe the sample used to train The Cannon, conduct cross-validation (\ref{sec:cross_validation}), and verify that the model is actually learning information about the asterosesimic parameter \Dnu\ and stellar age (\ref{sec:verify}). In Section \ref{sec:results}, we describe our results, including the table of inferred labels for our \textit{Kepler}-APOGEE test set (\ref{sec:kepler_test_sample}) and full APOGEE test set (\ref{sec:full_test_sample}) and their respective uncertainties (\ref{sec:uncertainties}). We also compare our results to an asteroseismic \textit{Kepler} gold standard sample (\ref{sec:legacy}). In Section \ref{sec:plato}, we discuss how this study can be applied to upcoming PLATO asteroseismic observations of FGK dwarf stars. Finally, we conclude in Section \ref{sec:conclusion}.

\section{The Cannon}
\label{sec:the-cannon}

The Cannon \citep{ness_cannon_2015, ho_label_2017} is a data-driven framework for predicting stellar parameters from stellar spectra. In this framework, a model is first trained on a set of labeled stars with well-characterized parameters and spectra. With this trained model in hand, as long as a star has spectra, we can use the model to infer its parameters (otherwise known as ``labels") for this star. For this work, we use The Cannon 2 \citep{casey_cannon_2016}, which features faster optimization and more complex flux modeling than the original software. The Cannon is comprised of a ``training phase" (during which the regression model is trained on training labels and spectra) and a ``test phase" (during which the trained model takes in spectra from the test sample in order to predict labels for those stars). 

\subsection{Training phase} \label{sec:training_phase}

The Cannon flux model can be described as:

\begin{equation}
    f_{\lambda, n} = v(l_n) \cdot \theta_\lambda + \textrm{noise},
\end{equation}

\noindent where $l_n$ is the vector of labels for each star, $v$ is the vectorizer, and $\theta_\lambda$ is the vector of coefficients at each pixel. The vectorizer can be as simple as $\sum{l_n}$, but for this work we find that linear terms alone are insufficient, so we choose $v$ to be a second-order polynomial expansion of $l_n$. The noise term is drawn from a Gaussian centered at zero with variance of $\sigma_{\lambda,n}^2 + s_\lambda^2$, where $\sigma_{\lambda,n}$ is the pixel-wise, per-star flux uncertainty for each training set spectrum and $s_\lambda^2$ is the model intrinsic pixel-wise scatter (this is reported by The Cannon upon training the model as \texttt{s2}). $\theta_{\lambda}$ is fit such that the following log-likelihood function is optimized:

\begin{equation}
    \textrm{ln}\hspace{0.5mm}p(f_{\lambda,n} | \theta_{\lambda}, l_n, {s_{\lambda}^2}) = - \frac{(f_{\lambda,n} - v(l_n) \cdot \theta_{\lambda})^2}{\sigma_{\lambda,n}^2 + s_\lambda^2} - \textrm{ln}(\sigma_{\lambda,n}^2 + s_\lambda^2)
\end{equation}.

\noindent Thus, at each wavelength pixel $\lambda$, the best-fit model coefficients and intrinsic scatter are those that maximize the likelihood:

\begin{equation}
    \theta_{\lambda}, s_{\lambda}^2 \leftarrow \underset{\theta,s}{\textrm{argmax}} [\sum_{n=0}^{N-1} \textrm{ln}\hspace{0.5mm}p(f_{\lambda,n} | \theta_{\lambda}, l_n, {s_{\lambda}^2})].
\end{equation}

Our model employs six labels that comprise $l_n$: \teff, \logg, [Fe/H], [Mg/H], age, and \Dnu. We find that $\nu_{max}$ contributes relatively little new information compared to the complexity it adds. Therefore, we omit it from our label space. Since we use a quadratic model in this work, this results in 28 features (one scalar normalization feature, six linear features, six squared features, and 15 cross-terms). For this study, in which our label space consists of only two elemental abundances, we opt not to add regularization to the model. 

\subsection{Test phase} \label{sec:test_phase}
After the model is trained, the optimized $\theta_{\lambda}$ and $s_{\lambda}^2$ are used to infer the labels $l_n$ of the test sample. The labels are predicted by taking the minimum of the negative log-likelihood functions, as follows: 

\begin{align}
\begin{split}
    l_n &= \underset{l_n}{\textrm{argmin}}[\sum_{\lambda=0}^{\Lambda-1}-\textrm{ln}\hspace{0.5mm}p(f_{\lambda n}|\theta_{\lambda}, l_n, s_{\lambda}^2)] \\
    & = \underset{l_n}{\textrm{argmin}}[\sum_{\lambda=0}^{\Lambda-1}-\frac{(f_{\lambda,n} - v(l_n) \cdot \theta_{\lambda})^2}{\sigma_{\lambda,n}^2 + s_\lambda^2}].
\end{split}
\end{align}

\noindent $l_n$ can then be fed back into the trained model to produce an inferred flux model for each wavelength pixel $\lambda$, and each stellar spectrum $n$. 

\section{Training Sample}
\label{sec:training_sample}
We construct our training set from the APOKASC catalog of FGK dwarf stars from \cite{serenelli_first_2017}, hereafter S17. The APOKASC catalog is comprised of stars observed with \textit{Kepler} and analyzed by the Kepler Asteroseismology Science Consortium (KASC), as well as with the Apache Point Observatory Galactic Evolution Experiment (APOGEE) spectrograph, which provides high-resolution ($R$ $\sim$ 22,500) spectra in the $H$-band (1.51$-$1.7 $\mu$m). The stellar parameters used in S17 came from the Sloan Digital Sky Survey (SDSS-IV)/APOGEE Data Release 13 (DR13; \citealt{albareti_13th_2017}), but in the era of SDSS-V/Milky Way Mapper (MWM; \citealt{bowen_optical_1973, gunn_25_2006, smee_multi-object_2013, wilson_apache_2019, kollmeier_sloan_2025}), we use updated \teff, \logg, and chemical abundances from SDSS-V Data Release 19 (DR19; \citealt{meszaros_sdss-v_2025}). For our training spectra, we use stacked spectra from the APOGEE spectra provided under the Astra framework in SDSS-V/MWM. 

After removing 7 stars from the S17 catalog with reported ages of -99 Gyr (denoting stars for which an age was unable to be obtained) and another 8 stars without APOGEE spectra from SDSS-V/MWM DR19, we have exactly 400 stars remaining. We then remove stars that are at the base of the red giant branch (RGB) based on the SDSS-V/MWM APOGEE Stellar Parameter and Chemical Abundances Pipeline (ASPCAP; \citealt{meszaros_sdss-v_2025}) \teff\ and \logg, in order to better scope our model for age inference of main sequence Sun-like stars. After excluding these early RGB stars (specifically, stars with \logg\ $<$ 3.8 and \teff\ $<$ 5250 K), we have 335 stars remaining in our training sample. 

We next apply continuum normalization on our training set spectra. As done in \cite{behmard_data-driven_2025}, we begin by removing the gaps between the APOGEE chips at 15790$-$15870 \r{A} and 16420$-$16490 \r{A}. We then compute the normalized flux as:

\begin{equation}
    \overline{f}(\lambda_0) = \frac{\sum_\lambda(f_\lambda \sigma_\lambda^{-2} w_\lambda(\lambda_0))}{\sigma_\lambda^{-2} w_\lambda(\lambda_0))},
\end{equation}

\noindent where $\lambda_0$ is the starting point wavelength of the smoothing window, $\sigma_\lambda$ is the flux uncertainty, and the weights $w_\lambda$ are:

\begin{equation}
    \mathrm{exp}\left[-{\frac{(\lambda_0-\lambda)^2}{2L^2}}\right]
\end{equation}

\noindent with the Gaussian smoothing window, $L$, set to 10 \r{A}. In this manner, weights $w_\lambda$ are calculated for every wavelength pixel in each spectrum. We set the flux errors to be very large (aka 999) at all bad pixel values (where the flux or flux uncertainties are zero or not finite). Finally, we clip all flux uncertainty values to a minimum of 0.005. This is common among SNR $>$ 200/pix APOGEE targets, whose flux uncertainties are underestimated \citep{mckinnon_data-driven_2024, behmard_data-driven_2025}\footnote{\url{For more information, see https://www.sdss4.org/dr16/irspec/spectra/}}.

\subsection{Cross-validation}
\label{sec:cross_validation}
To validate our training set, we perform k-fold cross-validation. We choose the number of folds as $k$ = 4. In doing so, we split the sample into four nearly equal-sized folds (three folds of 84 stars and one fold of 83 stars), and rotate each fold once as the test set and three times as part of the training set. For each of the four test sets, we independently train one model. For each star in a given test set, its corresponding model (trained on the other three-quarters of sample) is applied to the star's ASPCAP (\teff, \logg, [Fe/H], and [Mg/H]) and S17 (age and \Dnu) labels in order to produce a model spectrum. For each star, then, we can compute the $\chi^2$ as 

\begin{equation}
    \chi^2 = \sum{\frac{(f_{\lambda} - \overline{f}_{\lambda})^2}{\sigma_{\lambda}^2 + s_\lambda^2}},
\end{equation}

\noindent where $f_\lambda$ is the model spectrum and $\overline{f}_\lambda$ is the continuum-normalized APOGEE spectrum. 

From this first pass of 4-fold cross-validation, we notice high $\chi^2$ for several stars in our sample and manually vet their spectra. We uncover five multi-lined spectroscopic binaries: four SB2s (KIC 9025370, KIC 9693187, KIC 6145937, and KIC 9787965) and one SB3 (KIC 10610027). Additionally, we identify two stars with abnormally poor continuum normalization (KIC 7976303 and KIC 10454113). For both stars, poor skyline subtraction results in visually apparent skylines around 1524 nm. 
We removed these seven stars from our training set in order to ensure that the trained model does not bear the imprint of double- or triple-lines (or the missed skylines). We checked the remaining highest $\sim$15\% $\chi^2$ stars for fast rotators, whose higher magnetic activity might be responsible for smearing the spectra and inflating the $\chi^2$, and found none. 

\begin{figure*}
  {\begin{tabular}{@{}ccc@{}}
\includegraphics[width=0.30\textwidth]{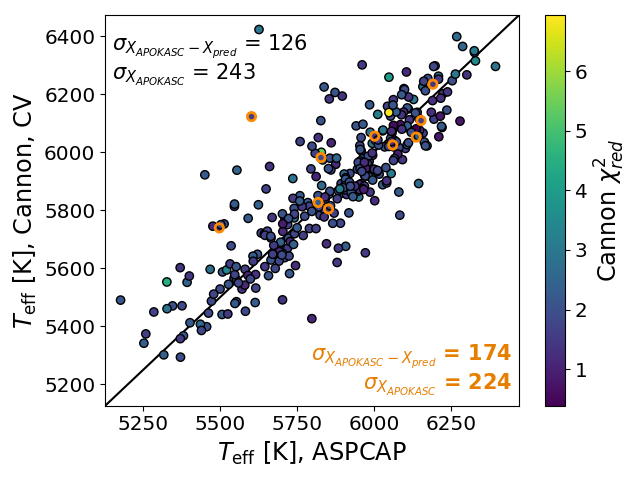}&
\includegraphics[width=0.30\textwidth]{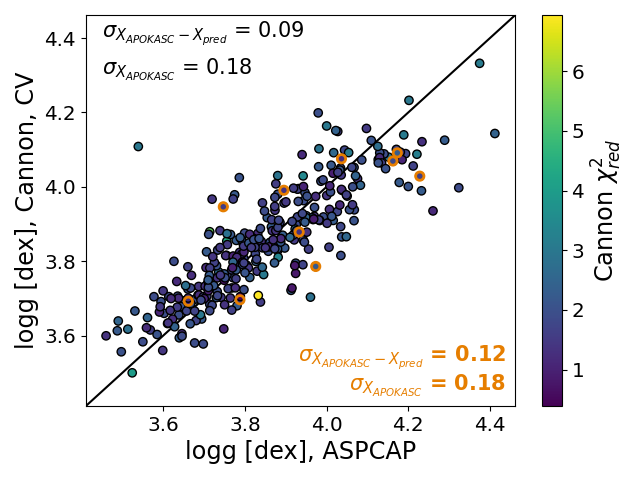}&
\includegraphics[width=0.30\textwidth]{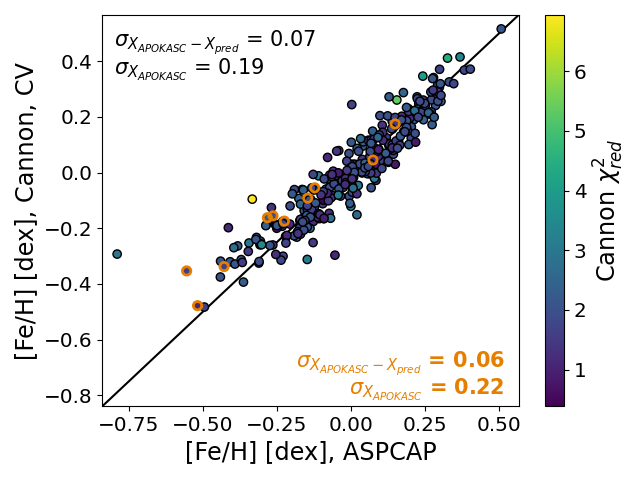}\\
\includegraphics[width=0.30\textwidth]{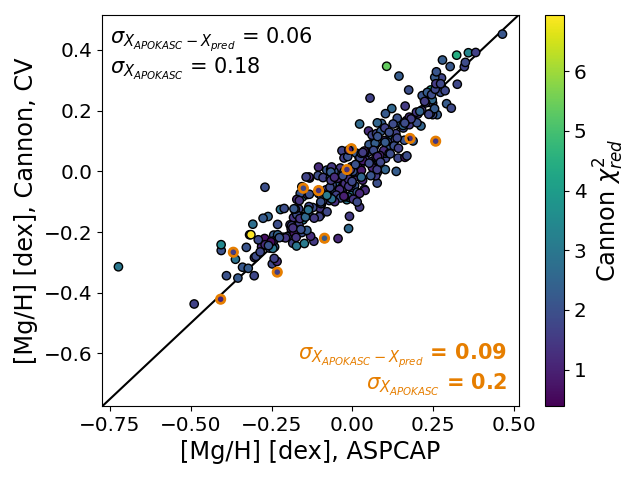}&
\includegraphics[width=0.30\textwidth]{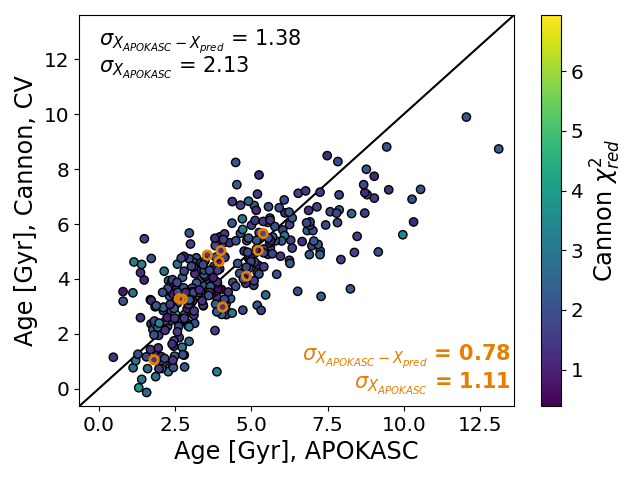}&
\includegraphics[width=0.30\textwidth]{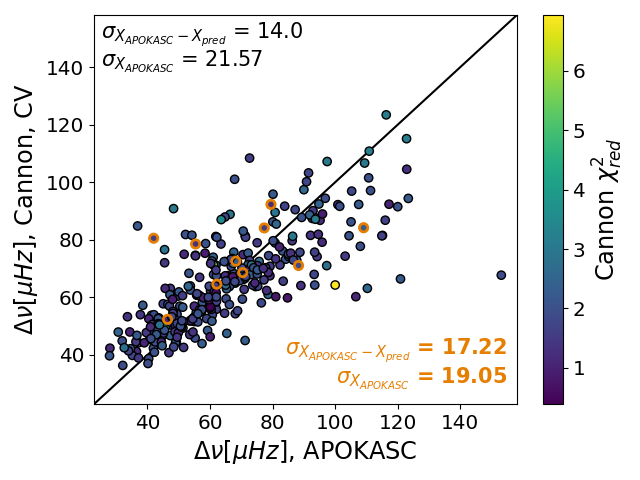}
  \end{tabular}}
    \caption{Predicted labels from The Cannon model, plotted against ASPCAP DR19 (for \teff, \logg, [Fe/H], and [Mg/H]) and S17 APOKASC (for age and \Dnu) labels and colored by $\chi^2$. The top left line in each box denotes the RMS scatter between the ASPCAP or APOKASC label and the corresponding label as predicted by The Cannon. The line beneath denotes the intrinsic ASPCAP/APOKASC label scatter. Young, alpha-rich stars ($\leq$6 Gyr, [Mg/Fe] $\geq$ 0.1 dex) are circled in orange, and the corresponding RMS scatter and intrinsic ASPCAP/APOKASC scatter are denoted in the bottom two lines on the right of each box. While there are some stellar ages for which The Cannon and ASPCAP/APOKASC disagree, our age inferences generally follow a 1:1 trend, with an RMS scatter of 1.38 Gyr.}
  \label{fig:loocv}
\end{figure*}

After removing the putative SB2s and SB3, as well as the two stars with poor skyline subtraction, we are left with 328 stars in our final training sample. Our training set is provided in Table \ref{tab:table1}. We perform 4-fold cross-validation once more on this final training set and report the results in Table \ref{tab:table2}. We compare these predictions, which represent the moderately degraded capabilities of The Cannon (because the model is trained on 75\% rather than 100\% of the training set), to their corresponding ASPCAP and S17 values in Figure \ref{fig:loocv}. In each box in Figure \ref{fig:loocv}, we note the RMS scatter between the APOKASC-derived labels and the Cannon-derived labels, as well as the intrinsic scatter among the APOKASC label. If the former is less than the latter, it means that there is a 1:1 correlation between the two label values that cannot be explained by the intrinsic scatter in the training set. We demonstrate that we stay under this intrinsic scatter for all parameters. However, we note that our age and \Dnu\ predictions are poorer for older stars ($\gtrsim$8 Gyr) with larger \Dnu\ values and systematically pulls them to smaller values. This is because our training set includes relatively fewer stars with these properties. We anticipate that more stars with these properties will be provided by future asteroseismology surveys (e.g., PLATO). 

Since this study is largely motivated by the need for stellar ages at scale, we compute the relative age error from four-fold cross-validation for each star in our training set. We report the median relative fractional age error to be 20.0\%. That is, for the typical star in our training set, its predicted age from The Cannon was 20\% away from the age measured by APOKASC.


\begin{table*}[]
\centering
    \begin{tabular}{c|c|c|c|c|c|c}
    KIC & \teff\ [K] & \logg \hspace{1mm}[dex] & [Fe/H] [dex] & [Mg/H] [dex] & Age [Gyr] & \Dnu\ [$\mu$Hz]\\
    \hline
    \hline
1435467 & 6111 & 3.83 & 0.08 & 0.01 & 2.60 & 70.56 \\
2010607 & 6102 & 3.57 & 0.06 & -0.00 & 2.75 & 42.48 \\
2450729 & 5825 & 3.90 & -0.23 & -0.18 & 6.03 & 61.91 \\
2837475 & 6209 & 3.74 & -0.03 & -0.06 & 1.13 & 75.72 \\
2849125 & 5959 & 3.51 & 0.30 & 0.22 & 2.58 & 41.32 \\
2852862 & 6058 & 3.63 & -0.08 & -0.10 & 2.26 & 53.56 \\
2991448 & 5615 & 3.97 & -0.09 & -0.04 & 6.88 & 61.73 \\
2998253 & 6108 & 4.03 & 0.08 & 0.01 & 2.62 & 87.23 \\
3112152 & 5946 & 3.84 & 0.00 & -0.00 & 3.78 & 65.13 \\
3112889 & 6035 & 3.77 & -0.23 & -0.24 & 3.63 & 53.71 
\end{tabular}
    \caption{Training set for this study. We combine the ages and \Dnu\ from APOKASC \citep{serenelli_first_2017} and \teff, \logg, [Fe/H], and [Mg/H] from ASPCAP \citep{meszaros_sdss-v_2025}. We show just the first ten rows to demonstrate the table's form and content, but a machine-readable version of the full training set is available. }
    \label{tab:table1}
\end{table*}

\begin{table*}[]
    \centering
    \begin{tabular}{c|c|c|c|c|c|c}
    Metric & \teff\ [K] & \logg \hspace{1mm}[dex] & [Fe/H] [dex] & [Mg/H] [dex] & Age [Gyr] & \Dnu\ [$\mu$Hz]\\
    \hline
    \hline
       RMSE & 126$\pm$19 & 0.11$\pm$0.02 & 0.06$\pm$0.00 & 0.06$\pm$0.00 & 1.41$\pm$0.16 & 14.7$\pm$1.3\\
       $\mu$ & 5889$\pm$14 & 3.85$\pm$0.01 & -0.00$\pm$0.02 & -0.02$\pm$0.02 & 4.29$\pm$0.09 & 65.6$\pm$0.56\\
       $\sigma$ & 219$\pm$14 & 0.13$\pm$0.01 & 0.17$\pm$0.00 & 0.16$\pm$0.01 & 1.80$\pm$0.01 & 14.0$\pm$1.0 \\
    \end{tabular}
    \caption{Results of 4-fold cross-validation, aggregated across all folds, after removal of lower red giant stars, SB3 and SB2s, and two cases of poor continuum normalization. Errors for each summary statistic are computed as the mean across all four cross-validation folds.}
    \label{tab:table2}
\end{table*}

\subsection{Age information validation} \label{sec:verify}
We have demonstrated that The Cannon can predict labels (including ages) of FGK dwarfs in a self-consistent manner. However, we must also show that these age predictions are not merely driven by the relation between Fe and Mg abundances and age, independent of the asteroseismic parameter, \Dnu. On average, younger stars tend to be $\alpha$-poor, while older stars tend to be more $\alpha$-rich \citep{haywood_age_2013, lu_spectroscopic_2026}. Therefore, our first test is to isolate only the young, $\alpha$-\textit{rich} stars from our training sample and check that our model is learning information from the input spectra, rather than simply following the expected broad trend of increasing $\alpha$ enhancement with increasing stellar age. We define young, $\alpha$-rich stars as age $\leq$ 6 Gyr and [Mg/Fe] $\geq$ 0.1 \citep{martig_young_2015, zhang_most_2021, maas_exploring_2025}, where [Mg/Fe] = [Mg/H] $-$ [Fe/H] (since chemical abundances in dex units correspond to log-space). In Figure \ref{fig:loocv}, we draw orange circles around these young, $\alpha$-rich stars and demonstrate that they follow the same 1:1 relation as the rest of the training set under 4-fold cross-validation. As with the broader cross-validation test, we show that even for this subset of our training sample, the intrinsic scatter for every APOKASC label is larger than the scatter of its RMS with respect to The Cannon's predicted labels. 

\begin{figure*}
  {\begin{tabular}{@{}ccc@{}}
\includegraphics[width=0.30\textwidth]{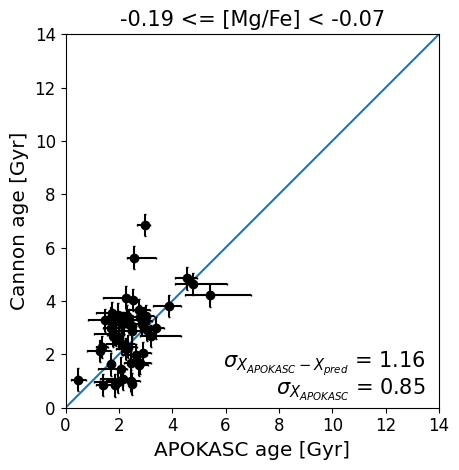}&
\includegraphics[width=0.30\textwidth]{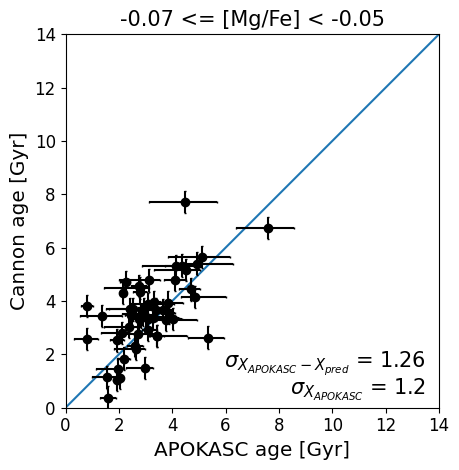}&
\includegraphics[width=0.30\textwidth]{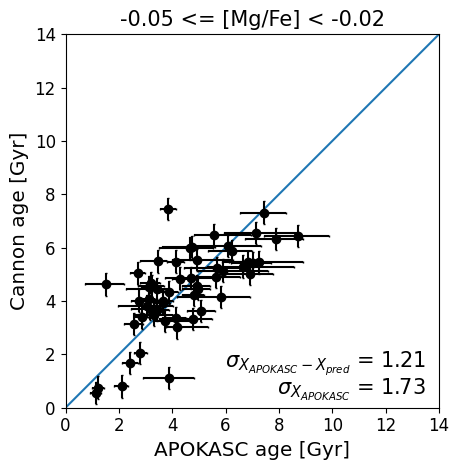}\\
\includegraphics[width=0.30\textwidth]{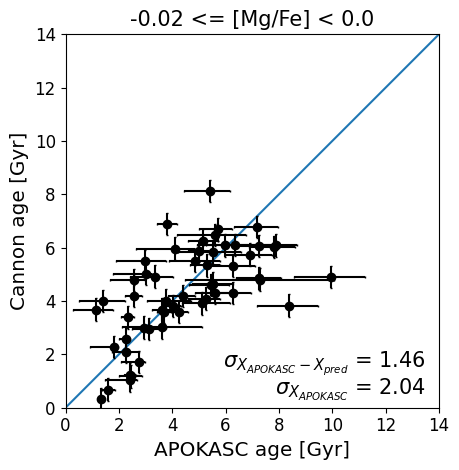}&
\includegraphics[width=0.30\textwidth]{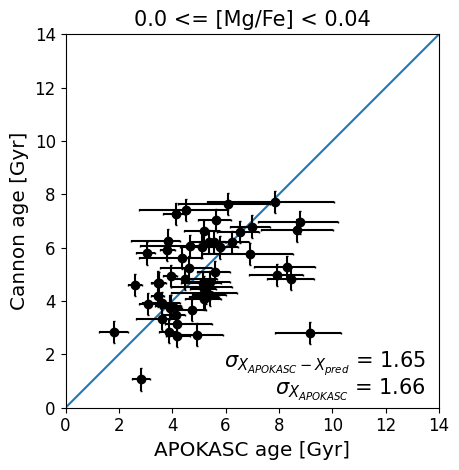}&
\includegraphics[width=0.30\textwidth]{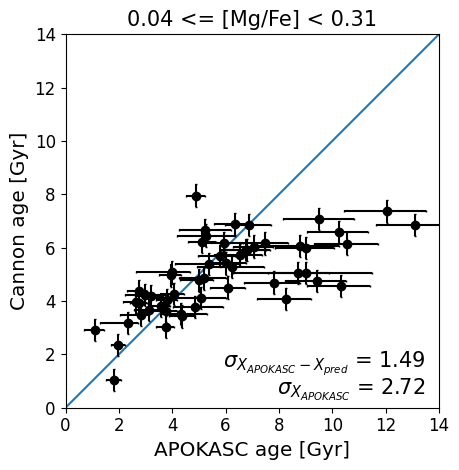}
  \end{tabular}}
    \caption{A mono-abundance analysis of our training sample in which we split the sample into six [Mg/Fe] bins with the same number of stars in each bin. If The Cannon's inferred ages were merely driven entirely by abundances, we should expect each bin to show no relation between The Cannon's age and the APOKASC age. Instead, for all but the lowest [Mg/Fe] bin, the 1:1 RMS scatter between the APOKASC and Cannon labels is smaller than the intrinsic scatter among the APOKASC labels for all bins, indicating that The Cannon's recovery of the age label is both accurate and informed by the spectra themselves. However, there is a bias, with The Cannon under-predicting ages of older stars.}
  \label{fig:mono-abundance-no-rgb-age}
\end{figure*}

We additionally apply a test commonly used in Galactic archaeology studies, in which the training sample is divided into six mono-abundance bins and then cross-validated. This is done to verify that our ability to predict the stellar age is not simply driven by learning the abundance trends encoded in the labels. If The Cannon is truly learning no age information outside of its dependence on iron and $\alpha$-abundances, then some bins will exhibit uniformly random scatter rather than a correlation between the Cannon and APOKASC parameters. As shown in Fig \ref{fig:mono-abundance-no-rgb-age}, however, this is not the case. We quantify this as done in Fig \ref{fig:loocv} by comparing the RMS scatter between the Cannon- and APOKASC-derived age for each mono-abundance bin to the intrinsic scatter for the corresponding APOKASC-derived parameter for that bin. For all mono-abundance bins, the RMS scatter is smaller than the intrinsic scatter, confirming that we are indeed learning asteroseismic information. As in 4-fold cross-validation, The Cannon under-predicts ages of older stars. This is most apparent for the highest [Mg/Fe] mono-abundance bins, although this is likely due to higher [Mg/Fe] bins containing more older ($\geq$6 Gyr) stars. 

\section{Results}
\label{sec:results}

\subsection{\textit{Kepler}-APOGEE test sample} \label{sec:kepler_test_sample}
Our test sample is comprised of all \textit{Kepler} stars with APOGEE spectra, culled to resemble the training set in ASPCAP \teff, \logg, [Fe/H], and [Mg/H] space. The initial test set of \textit{Kepler} stars with APOGEE spectra contains 6131 stars. We then remove stars that lie beyond the minimum or maximum \teff, \logg, [Fe/H], or [Mg/H] of our training sample, resulting in 3812 stars. We restrict our test sample to be contained within the \teff, \logg, [Fe/H], and [Mg/H] parameter space spanned by our training set, i.e., the minimum and maximum limits:
\begin{enumerate}
    \item 5176 $\leq$ \teff\ $\leq$ 6395 K
    \item 3.46 $\leq$ \logg\ $\leq$ 4.41 dex
    \item $-$0.79 $\leq$ [Fe/H] $\leq$ 0.51 dex
    \item $-$0.72 $\leq$ [Mg/H] $\leq$ 0.46 dex
\end{enumerate}

\begin{figure*}
\centering
\includegraphics[width=0.85\textwidth]{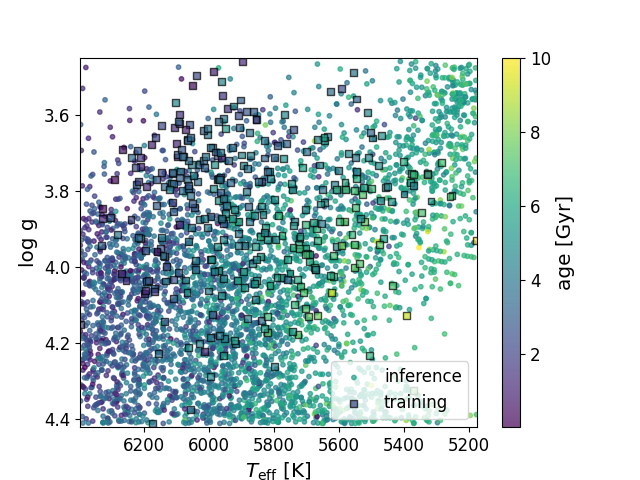}
\caption{Kiel diagram for our 3810 test sample (circles), with training set stars overlaid (squares), using ASPCAP labels. Stars are color-coded by the inferred age from our fully-trained model.} 
\label{fig:kiel-inference-color-coded}
\end{figure*}

We make an additional cut to the test set based on the model spectral fit $\chi^2$, which is calculated for each test star by generating a model spectrum using the inferred labels and comparing it to the input continuum-normalized spectrum. We omit stars where the $\chi^2$ is greater than 100,000 \citep{behmard_data-driven_2025, birky_temperatures_2020}, which corresponds to a reduced $\chi^2$ threshold of approximately 13.5 (with 7409 pixels and 6 fitted parameters, there are 7403 degrees of freedom). This removes two more stars from our sample, KIC 3341831 and KIC 6467256, leaving a final catalog of 3810 stars. Table \ref{tab:table3} reports the label value (and uncertainties, see next section) from The Cannon for each star in the final test sample, along with the reduced $\chi^2$. In Figure \ref{fig:kiel-inference-color-coded}, we show the test sample (circles) overlaid by the training sample (squares) in ASPCAP \teff\ and \logg\ space, color-coded by stellar ages as predicted by The Cannon. 

\begin{table*}[]
\centering
\begin{tabular}{rrllllllr}
    KIC & SDSS ID & \teff\ [K] & \logg \hspace{1mm}[dex] & [Fe/H] [dex] & [Mg/H] [dex] & Age [Gyr] & \Dnu\ [$\mu$Hz] & $\chi_{red}^2$\\
    \hline
    \hline
12689153 & 67743667 & 5776 $\pm$ 31 & 4.17 $\pm$ 0.04 & -0.04 $\pm$ 0.03 & -0.03 $\pm$ 0.03 & 5.60 $\pm$ 0.42 & 116.2 $\pm$ 0.8 & 1.45 \\
12735580 & 67800894 & 5894 $\pm$ 31 & 3.90 $\pm$ 0.05 & 0.17 $\pm$ 0.03 & 0.08 $\pm$ 0.03 & 3.90 $\pm$ 0.42 & 70.2 $\pm$ 0.8 & 2.15 \\
12735793 & 67800901 & 5843 $\pm$ 31 & 4.09 $\pm$ 0.04 & -0.10 $\pm$ 0.03 & -0.10 $\pm$ 0.03 & 5.47 $\pm$ 0.42 & 99.7 $\pm$ 0.8 & 1.59 \\
12736465 & 67800824 & 5971 $\pm$ 31 & 3.90 $\pm$ 0.06 & -0.03 $\pm$ 0.03 & -0.11 $\pm$ 0.03 & 3.41 $\pm$ 0.42 & 71.7 $\pm$ 0.8 & 1.21 \\
12737160 & 67800533 & 5760 $\pm$ 31 & 4.02 $\pm$ 0.04 & 0.17 $\pm$ 0.03 & 0.10 $\pm$ 0.03 & 5.09 $\pm$ 0.42 & 71.6 $\pm$ 0.8 & 1.96 \\
12783625 & 67800923 & 6035 $\pm$ 31 & 3.98 $\pm$ 0.05 & 0.08 $\pm$ 0.03 & -0.02 $\pm$ 0.03 & 2.91 $\pm$ 0.42 & 80.0 $\pm$ 0.8 & 1.72 \\
12784610 & 67800812 & 5684 $\pm$ 31 & 4.08 $\pm$ 0.04 & -0.04 $\pm$ 0.03 & -0.04 $\pm$ 0.03 & 6.34 $\pm$ 0.42 & 105.0 $\pm$ 0.8 & 2.05 \\
12833827 & 67800943 & 5309 $\pm$ 31 & 3.89 $\pm$ 0.04 & -0.08 $\pm$ 0.03 & 0.00 $\pm$ 0.03 & 7.99 $\pm$ 0.42 & 54.9 $\pm$ 0.8 & 2.02 \\
12834387 & 67800878 & 6251 $\pm$ 31 & 3.92 $\pm$ 0.05 & -0.00 $\pm$ 0.03 & -0.05 $\pm$ 0.03 & 1.20 $\pm$ 0.42 & 86.1 $\pm$ 0.8 & 1.82 \\
12834614 & 67801064 & 6138 $\pm$ 31 & 3.62 $\pm$ 0.05 & -0.01 $\pm$ 0.03 & -0.06 $\pm$ 0.03 & 1.70 $\pm$ 0.42 & 50.4 $\pm$ 0.8 & 1.40 \\
\end{tabular}
    \caption{Test sample for this study, with predicted parameters for each of the 3810 stars and reduced $\chi^2$ computed from the input spectrum and the spectrum generated by the trained model. The test sample is constructed from stars in the KIC with ASPCAP parameters within the max and min bounds of the training sample in \teff, logg, [Fe/H], and [Mg/H] space. Inferred parameter errors are similar from star to star because the bulk of the error budget is driven by the global $\sigma_{\mathrm{inflate}}$ rather than the star-to-star model systematic error (see Section \ref{sec:uncertainties}). We show just the first ten rows to demonstrate the form and content of this table. A machine-readable version of the full table is available. }
    \label{tab:table3}
\end{table*}

\subsection{Full APOGEE test sample} \label{sec:full_test_sample}
The test sample can be further expanded to include all stars visited by APOGEE within our training label space, regardless of membership in the \textit{Kepler} Input Catalog (KIC). After removing APOGEE stars with SNR$\leq$200 and whose inferred spectra from The Cannon have a $\chi^2$$\geq$100,000, we are left with an expanded test set of 48,828. These stars' predicted \teff, \logg, [Fe/H], [Mg/H], age, and \Dnu, along with the reduced $\chi^2$ of their inferred spectra, are shown in Table \ref{tab:table4}.

\begin{table*}[]
\centering
\begin{tabular}{rrllllllr}
\textit{Gaia} DR3 Source ID & SDSS ID & \teff\ [K] & \logg\ [dex] & [Fe/H] [dex] & [Mg/H] [dex] & Age [Gyr] & \Dnu\ [$\mu$Hz] & $\chi_{red}^2$ \\
\hline
\hline
420868586978944768 & 117524371 & 5943 $\pm$ 31 & 4.04 $\pm$ 0.04 & -0.02 $\pm$ 0.03 & -0.00 $\pm$ 0.03 & 3.77 $\pm$ 0.42 & 85.2 $\pm$ 0.8 & 2.06 \\
421036129367476352 & 117526412 & 6016 $\pm$ 31 & 3.87 $\pm$ 0.04 & -0.33 $\pm$ 0.03 & -0.34 $\pm$ 0.03 & 4.69 $\pm$ 0.42 & 61.4 $\pm$ 0.8 & 2.31 \\
421238332125586816 & 117528719 & 6027 $\pm$ 31 & 3.79 $\pm$ 0.05 & 0.06 $\pm$ 0.03 & -0.02 $\pm$ 0.03 & 2.93 $\pm$ 0.42 & 58.1 $\pm$ 0.8 & 1.77 \\
4111233547185089536 & 129044157 & 5584 $\pm$ 31 & 4.13 $\pm$ 0.04 & -0.13 $\pm$ 0.03 & 0.04 $\pm$ 0.03 & 8.60 $\pm$ 0.42 & 78.9 $\pm$ 0.8 & 2.67 \\
5337389398147162112 & 129049943 & 6255 $\pm$ 31 & 3.98 $\pm$ 0.04 & -0.15 $\pm$ 0.03 & -0.25 $\pm$ 0.03 & 2.83 $\pm$ 0.42 & 84.7 $\pm$ 0.8 & 3.37 \\
2075118195014471680 & 129051766 & 6102 $\pm$ 31 & 3.81 $\pm$ 0.05 & -0.07 $\pm$ 0.03 & -0.15 $\pm$ 0.03 & 3.20 $\pm$ 0.42 & 69.6 $\pm$ 0.8 & 1.83 \\
2101443973710952448 & 129055079 & 5827 $\pm$ 31 & 3.81 $\pm$ 0.04 & -0.13 $\pm$ 0.03 & -0.19 $\pm$ 0.03 & 3.47 $\pm$ 0.42 & 49.5 $\pm$ 0.8 & 2.23 \\
120903986512700160 & 129058214 & 5769 $\pm$ 31 & 3.85 $\pm$ 0.05 & 0.03 $\pm$ 0.03 & 0.02 $\pm$ 0.03 & 5.67 $\pm$ 0.42 & 64.5 $\pm$ 0.8 & 1.53 \\
6036578840946969600 & 129058425 & 6387 $\pm$ 31 & 3.99 $\pm$ 0.04 & -0.45 $\pm$ 0.03 & -0.53 $\pm$ 0.03 & 1.28 $\pm$ 0.42 & 66.5 $\pm$ 0.8 & 3.11 \\
120953464536688128 & 129059538 & 6064 $\pm$ 31 & 4.00 $\pm$ 0.05 & -0.24 $\pm$ 0.03 & -0.23 $\pm$ 0.03 & 4.73 $\pm$ 0.42 & 78.1 $\pm$ 0.8 & 1.48 \\
\end{tabular}
    \caption{Expanded test sample, with predicted parameters for each of the 48,828 stars and reduced $\chi^2$ computed from the input spectrum and the spectrum generated by the trained model. The test sample is constructed from stars with ASPCAP parameters within the max and min bounds of the training sample in \teff, logg, [Fe/H], and [Mg/H] space. Inferred parameter errors are similar from star to star because the bulk of the error budget is driven by the global $\sigma_{\mathrm{inflate}}$ rather than the star-to-star model systematic error (see Section \ref{sec:uncertainties}). We show just the first ten rows to demonstrate the form and content of this table. A machine-readable version of the full table is available. }
    \label{tab:table4}
\end{table*}

\subsection{Predicted label uncertainties}
\label{sec:uncertainties}

In order to quantify the uncertainties on our inferred test set labels, we employ the procedure used in \citep{behmard_data-driven_2025}, in which the inter-observation scatter introduced by different APOGEE observations is estimated for 500 stars with multiple APOGEE visits that lie within the training sample parameter space. Together with the intrinsic model scatter from The Cannon, this observational uncertainty amounts to the overall uncertainty per parameter per star in our test sample. While the model uncertainties for each of the two spectra $A$ and $B$, $\sigma_A$ and $\sigma_B$, are readily provided by the covariance matrices from The Cannon, we must fit the observational uncertainty using the following equation for each parameter:
\begin{equation}
    Z_{A,B} = \frac{l_A - l_B}{\sqrt{\sigma_A^2 + \sigma_B^2 + 2(\sigma_{\mathrm{inflate}}^2)}} \hspace{1mm},
\end{equation}

where $l_A$ and $l_B$ are the inferred label for each of the two spectra $A$ and $B$, $\sigma_A$ and $\sigma_B$ are the scatter from The Cannon model estimated during the test phase, and $\sigma_{\mathrm{inflate}}$ is the inter-observation scatter that we are trying to fit, such that $Z_{A,B}$ for the 500 stars follows a unit normal distribution. We therefore fit for a $\sigma_{\mathrm{inflate}}$ for each label, and report the test sample uncertainties in Table \ref{tab:table3} as the quadrature sum of $\sigma_{\mathrm{inflate}}$ and the scatter for each star from the model covariance matrix (analogous to $\sigma_A$ and $\sigma_B$). As in \citealt{behmard_data-driven_2025}, the global $\sigma_{\mathrm{inflate}}$ contributes the bulk of the total error budget, compared to the Cannon model error, which varies from star to star. 

One implication of a nearly constant age uncertainty is that the fractional age uncertainty varies almost entirely directly with age. While the majority (72\%) of stars in our test sample have age predictions under 10\% fractional age uncertainty, stars younger than 4.2 Gyr will by construction have fractional age uncertainties larger than 10\% (and stars younger than 2.1 Gyr will have fractional age uncertainties larger than 20\%, and so on). However, we should stress that a better measure of this method's efficacy at age inference is the relative fractional age \textit{error} from cross-validation compared to the APOKASC ages, which was reported to be 20\% in Section \ref{sec:cross_validation}.

\subsection{Kepler LEGACY sample}
\label{sec:legacy}
\begin{figure*}
\centering
\includegraphics[width=.7\textwidth]{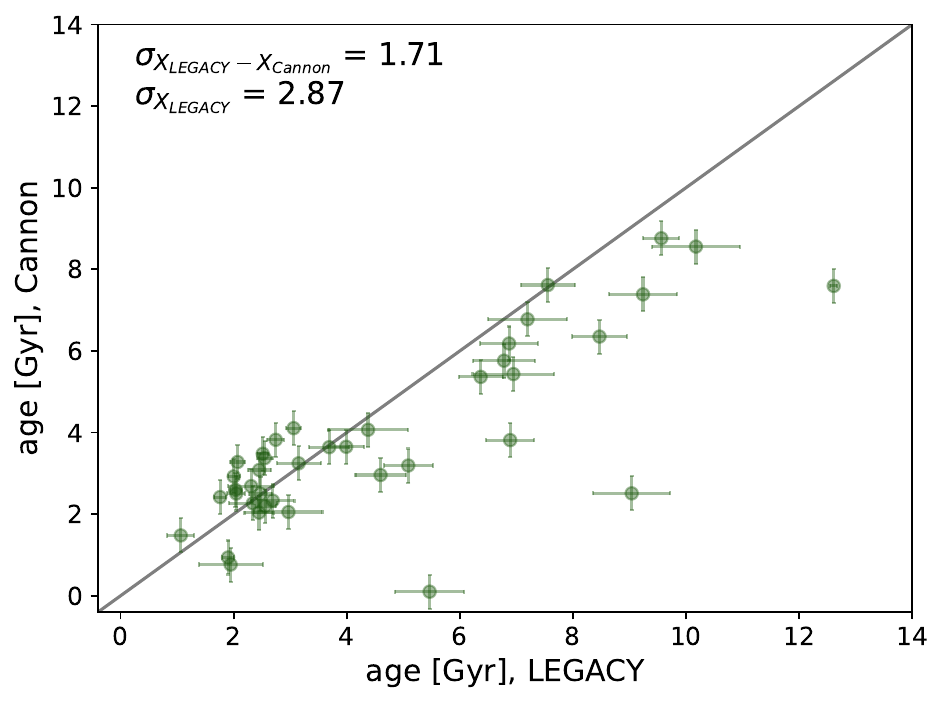}
\caption{A comparison between stellar ages predicted by The Cannon for the \textit{Kepler} asteroseismic LEGACY sample \citep{aguirre_standing_2017}. The RMS scatter between the Cannon and LEGACY ages outperforms the intrinsic scatter from LEGACY. However, as in cross-validation, our model tends under-predict ages of older stars.}
\label{fig:legacy}
\end{figure*}

So far, we have only used one sample of independently-derived asteroseismic parameters and ages for \textit{Kepler} stars \citep{serenelli_first_2017}. \cite{aguirre_standing_2017} presented a smaller but more precisely measured \textit{Kepler} asteroseismic sample called LEGACY. With only 66 stars, however, it is better suited for our purposes as a gold standard comparison sample rather than a training sample. After cross-matching the LEGACY sample with our test sample, we plot the ages of the remaining 40 stars in Figure \ref{fig:legacy} and show that, once again, the RMS scatter is smaller than the intrinsic scatter among the LEGACY ages. As previously shown in Figures \ref{fig:loocv} and \ref{fig:mono-abundance-no-rgb-age}, the Cannon tends to under-predict ages of older stars in the LEGACY sample. This can be explained by a lack of training data in that stellar age regime and can be improved by a larger training sample from future surveys such as PLATO, which will observe older stars in the thick disk \citep{boettner_exoplanets_2024, rauer_plato_2024}. Nevertheless, in general, together with satisfactory 4-fold cross-validation and mono-abundance test results, this demonstrates that The Cannon is robust at inferring the global asteroseismology parameter \Dnu\ and stellar ages for \textit{Kepler} FGK dwarfs younger than $\sim$6 Gyr.  

\section{Looking Ahead to the PLATO Data Release} \label{sec:plato}

Comprising 26 cameras sensitive to 13 $\geq$ $m_v$ $\geq$ 4, with a field of view of 49"x49" (4510x4510 pixels), ESA's PLATO will conduct a large asteroseismology survey of potential planet hosts in the Solar neighborhood and is scheduled to launch in January 2027 \citep{rauer_plato_2024}. With better asteroseismic parameters at scale comes better stellar ages at scale \citep{goupil_predicted_2024}, potentially enabling exoplanet demographic studies that rely on stellar ages to more exquisite precision. PLATO stands to provide a larger training sample with broader parameter space coverage than the S17 sample used in this work, enabling more precise stellar age inference for a larger label space of stars. 

\subsection{PLATO stars as an enhanced training set} \label{sec:plato_stars}

\begin{figure*}
\centering
\includegraphics[width=.7\textwidth]{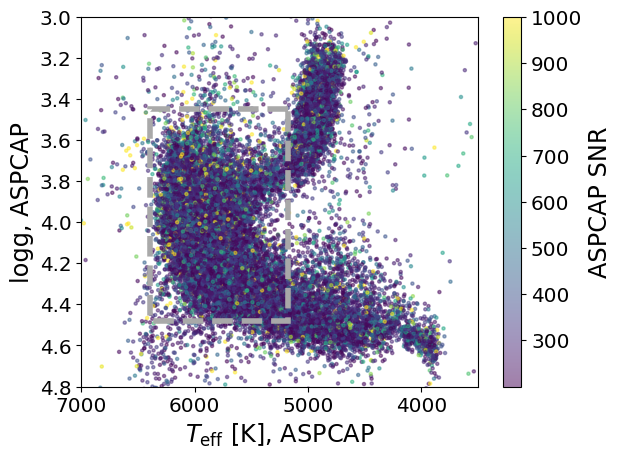}
\caption{Kiel diagram of the cross-match between the asPIC (v.1; \cite{montalto_all-sky_2021}) and stars with APOGEE spectra, limited to approximate solar-like dwarf and sub-giant space. There are 24,642 stars in the cross-match, and 23,772 of these are plotted here. The dashed box demarcates the label space from this work. PLATO will extend the training label space to cooler main sequence stars, enabling high-precision age inference for a stellar regime that is traditionally intractable to age date.}
\label{fig:kiel_plato_aspcap}
\end{figure*}

The all-sky PLATO Input Catalog (asPIC) \citep{montalto_all-sky_2021} is a catalog of all the stars that will be targeted by PLATO's southern and (if applicable) northern fields. It contains 2,378,177 FGK dwarfs, of which 64,105 also have APOGEE spectra through Data Release 19 of SDSS-V/MWM. Of these stars, we keep the 24,642 stars with APOGEE SNR $\geq$ 200, and of those stars, 23,772 lie within 3500 $\leq$ \teff\ $\leq$ 7000 K and 3.0 $\leq$ \logg $~\leq$ 4.8 dex, which are pictured in Figure \ref{fig:kiel_plato_aspcap}. In chemical abundance space, these stars span $-$2.49 $\leq$ [Fe/H] $\leq$ 1.01 dex and $-$2.43 $\leq$ [Mg/H] $\leq$ 1.81 dex. This sample would potentially be a new training set for The Cannon that is both two orders of magnitude larger than that used in this study and more comprehensive in its label space coverage (as shown in Figure \ref{fig:kiel_plato_aspcap}, we will be able to probe cooler main sequence stars thanks to PLATO's brighter targets). One could then construct a test sample by taking the subset of stars with APOGEE DR19 spectra that lie within this expanded label space, resulting in 437,767 stars for which asteroseismic-quality ages are possible, regardless of whether they will be observed by PLATO. 

\subsection{PLATO planets} \label{sec:plato_planets}
Stellar ages are vital for probing how exoplanet populations evolve. For example, they are needed for understanding planet occurrence rates (e.g., \citealt{sayeed_homogeneous_2025}, \citealt{lam_late-time_2025}), planet architecture sculpting (e.g., \citealt{lam_ages_2024}, \citealt{chen_planets_2021}), and how the planet radius gap changes over time (e.g., \citealt{vaneylen_asteroseismic_2018}, \citealt{berger_gaia-kepler_2020}, \citealt{weeks_asteroseismology_2025}, \citealt{chen_insights_2026}). The degree of precision to which stellar ages are measured in turn affects the degree to which models of exoplanet population evolution are discernible. For example, \citealt{lam_ages_2024} showed that models of Gyr-timescale dynamical sculpting become discernible at ages measured to better than 10\% precision. 

PLATO will provide uniformly- and well-measured asteroseismic ages for thousands to tens of thousands of planet hosts \citep{rauer_plato_2024, boettner_exoplanets_2024, matuszewski_estimating_2023}. Moreover, PLATO's overlap with the \textit{Kepler} \citep{nascimbeni_plato_2025} and TESS \citep{eschen_viewing_2024} fields will provide precise age measurements for a modest number of known planets. Stellar spectrum-driven label transfer of asteroseismic ages can leverage the confluence of PLATO, \textit{Kepler}, and TESS to expand the sample of planets with high quality ages to demographically relevant quantities.

From our test set, there are 1092 \textit{Kepler} confirmed planets\footnote{\url{https://exoplanetarchive.ipac.caltech.edu/docs/data.html}, accessed 8 July 2026.} for which we readily have Cannon ages, compared to 77 \textit{Kepler} planets in the PLATO input catalog, 16 \textit{Kepler} planets in the S17 sample, and 5 \textit{Kepler} planets in the LEGACY sample. In other words, even if PLATO were to observe the northern field for the third and fourth years of its mission, obtaining asteroseismic-quality ages for \textit{Kepler} planets hosted by main sequence stars at quantities suitable for age-based demographics is not possible without the help of tools like The Cannon. 

The proof of concept demonstrated in this study will also synergize with PLATO. From the expanded label space of the PLATO training sample (Figure \ref{fig:kiel_plato_aspcap}), there will be an additional 720 \textit{Kepler} confirmed planets with APOGEE spectra for which we will be able to determine asteroseismic-quality ages. Additionally, while there are very few Transiting Exoplanet Survey Satellite (TESS) planet hosts with asteroseismic ages (e.g., \citealt{nielsen_tess_2020}), of the 928 confirmed TESS planets\footnote{As of 8 July 2026.}, there will be 237 such planets with host stars in the expanded APOGEE label space after training on PLATO stars. In comparison, there are 134 confirmed TESS planets in the PLATO Input Catalog (including in the northern sky, which is not yet guaranteed to be observed) and 101 planets with APOGEE spectra in the current label space (without training on PLATO stars) for which we can readily obtain high-quality age inferences. 

\section{Conclusion} \label{sec:conclusion}

In this study, we trained The Cannon on the APOGEE DR19 spectra of 328 FGK dwarfs from the \cite{serenelli_first_2017} sample of solar-like stars observed by \textit{Kepler} with asteroseismology measurements, along with key spectroscopic (\teff, \logg, [Fe/H], and [Mg/H]; all from APOGEE DR19) and asteroseismic (age and \Dnu) parameters (which we refer to as ``labels" in our data-driven model). We validated our model using a variety of tests, including 4-fold cross-validation; isolating young, $\alpha-$rich stars; and taking mono-abundance slices of the training set. We then apply our model to APOGEE DR19 spectra and predict stellar ages for 48,828 FGK dwarfs in our training label space (3810 of which are also in the \textit{Kepler} Input Catalog). We compare our inferred ages with those of an independently measured asteroseismic age sample, \emph{Kepler} LEGACY \citep{aguirre_standing_2017}, and found that the stellar ages of the cross-match of these two samples are in good agreement. This is one of the first studies to extend stellar ages to larger samples for FGK dwarfs via data-driven techniques (other studies, which leveraged gyrochronological ages, include \citealt{sayeed_homogeneous_2025} and \citealt{van-lane_chronoflow_2025}).

The method demonstrated here will have dramatic utility with the upcoming PLATO data release \citep{rauer_plato_2014, montalto_all-sky_2021, rauer_plato_2024}. There are 64,105 stars in the asPIC$-$APOGEE DR19 cross-match, which occupy a larger label space than our current training set (notably, with better coverage of cooler stars), and 439,005 stars with APOGEE DR19 spectra that lie within this expanded label space. Within this potential test set are 1812 \textit{Kepler} confirmed planets and 165 TESS candidate planets. We will be able to infer ages for these planet host stars once we train The Cannon on the PLATO data release. This will enable population-level studies of planetary system changes over Gyr timescales \citep{boettner_exoplanets_2024}, thus robustly extending planet demographics studies along the age axis.

\section*{Acknowledgments}
We thank Adrian Price-Whelan, Andy Casey, Catherine Manea, Luke Bouma, Quadry Chance, and the CCA Astronomical Data, CCA Spectroscopy, and CCA Nearby Universe Groups for useful discussions that have helped inform and enrich this work. This material is based upon work supported by the National Science Foundation Graduate Research Fellowship Program under Grant No. 1842473.

Funding for the Sloan Digital Sky Survey V has been provided by the Alfred P. Sloan Foundation, the Heising-Simons Foundation, the National Science Foundation, and the Participating Institutions. SDSS acknowledges support and resources from the Center for High-Performance Computing at the University of Utah. SDSS telescopes are located at Apache Point Observatory, funded by the Astrophysical Research Consortium and operated by New Mexico State University, and at Las Campanas Observatory, operated by the Carnegie Institution for Science. The SDSS web site is \url{www.sdss.org}.

SDSS is managed by the Astrophysical Research Consortium for the Participating Institutions of the SDSS Collaboration, including the Carnegie Institution for Science, Chilean National Time Allocation Committee (CNTAC) ratified researchers, Caltech, the Gotham Participation Group, Harvard University, Heidelberg University, The Flatiron Institute, The Johns Hopkins University, L'Ecole polytechnique f\'{e}d\'{e}rale de Lausanne (EPFL), Leibniz-Institut f\"{u}r Astrophysik Potsdam (AIP), Max-Planck-Institut f\"{u}r Astronomie (MPIA Heidelberg), Max-Planck-Institut f\"{u}r Extraterrestrische Physik (MPE), Nanjing University, National Astronomical Observatories of China (NAOC), New Mexico State University, The Ohio State University, Pennsylvania State University, Smithsonian Astrophysical Observatory, Space Telescope Science Institute (STScI), the Stellar Astrophysics Participation Group, Universidad Nacional Aut\'{o}noma de M\'{e}xico, University of Arizona, University of Colorado Boulder, University of Illinois at Urbana-Champaign, University of Toronto, University of Utah, University of Virginia, Yale University, and Yunnan University.

We acknowledge that IPAC/Caltech resides on the traditional, ancestral and unceded territory of the Gabrielino/Tongva peoples, the original caretakers of this land. We respectfully recognize the Gabrielino/Tongva peoples who still reside in Tovaangar (the Los Angeles basin and South Channel Islands). We also acknowledge that the main campus of the University of Florida is located on the ancestral territory of the Potano and of the Seminole peoples. The Potano, of Timucua affiliation, lived here in the Alachua region from before European arrival until the destruction of their towns in the early 1700s. The Seminole, also known as the Alachua Seminole, established towns here shortly after but were forced from the land as a result of a series of wars with the United States known as the Seminole Wars. We, the authors, acknowledge our obligation to honor the past, present, and future Native residents and cultures of California and Florida. Finally, we acknowledge the land politically designated as New York City to be the homeland of the Lenape (Lenapehoking) who were violently displaced as a result of European settler colonialism over the course of 400 years. The Lenape are a diasporic people that remain closely connected with this land and are its rightful stewards. We also recognize that New York City has one of the largest urban Native American / Indigenous populations in the United States.

\bibliography{cannon-ages}{}

@article{serenelli_first_2017,
	title = {The {First} {APOKASC} {Catalog} of {Kepler} {Dwarf} and {Subgiant} {Stars}},
	volume = {233},
	issn = {0067-0049},
	url = {https://dx.doi.org/10.3847/1538-4365/aa97df},
	doi = {10.3847/1538-4365/aa97df},
	language = {en},
	number = {2},
	urldate = {2025-03-19},
	journal = {The Astrophysical Journal Supplement Series},
	publisher = {The American Astronomical Society},
	author = {Serenelli, Aldo and Johnson, Jennifer and Huber, Daniel and Pinsonneault, Marc and Ball, Warrick H. and Tayar, Jamie and Aguirre, Victor Silva and Basu, Sarbani and Troup, Nicholas and Hekker, Saskia and Kallinger, Thomas and Stello, Dennis and Davies, Guy R. and Lund, Mikkel N. and Mathur, Savita and Mosser, Benoit and Stassun, Keivan G. and Chaplin, William J. and Elsworth, Yvonne and García, Rafael A. and Handberg, Rasmus and Holtzman, Jon and Hearty, Fred and García-Hernández, D. A. and Gaulme, Patrick and Zamora, Olga},
	month = dec,
	year = {2017},
	pages = {23},
}

@article{ness_cannon_2015,
	title = {{THE} {CANNON}: {A} {DATA}-{DRIVEN} {APPROACH} {TO} {S}℡{LAR} {LABEL} {DETERMINATION}},
	volume = {808},
	issn = {0004-637X},
	shorttitle = {{THE} {CANNON}},
	url = {https://dx.doi.org/10.1088/0004-637X/808/1/16},
	doi = {10.1088/0004-637X/808/1/16},
	language = {en},
	number = {1},
	urldate = {2025-03-19},
	journal = {The Astrophysical Journal},
	publisher = {The American Astronomical Society},
	author = {Ness, M. and Hogg, David W. and Rix, H.-W. and Ho, Anna. Y. Q. and Zasowski, G.},
	month = jul,
	year = {2015},
	pages = {16},
}

@article{birky_temperatures_2020,
	title = {Temperatures and {Metallicities} of {M} {Dwarfs} in the {APOGEE} {Survey}},
	volume = {892},
	issn = {0004-637X},
	url = {https://dx.doi.org/10.3847/1538-4357/ab7004},
	doi = {10.3847/1538-4357/ab7004},
	language = {en},
	number = {1},
	urldate = {2025-05-08},
	journal = {The Astrophysical Journal},
	publisher = {The American Astronomical Society},
	author = {Birky, Jessica and Hogg, David W. and Mann, Andrew W. and Burgasser, Adam},
	month = mar,
	year = {2020},
	pages = {31},
}

@article{goupil_predicted_2024,
	title = {Predicted asteroseismic detection yield for solar-like oscillating stars with {PLATO}},
	volume = {683},
	copyright = {https://creativecommons.org/licenses/by/4.0},
	issn = {0004-6361, 1432-0746},
	url = {https://www.aanda.org/10.1051/0004-6361/202348111},
	doi = {10.1051/0004-6361/202348111},
	language = {en},
	urldate = {2025-05-08},
	journal = {Astronomy \& Astrophysics},
	author = {Goupil, M. J. and Catala, C. and Samadi, R. and Belkacem, K. and Ouazzani, R. M. and Reese, D. R. and Appourchaux, T. and Mathur, S. and Cabrera, J. and Börner, A. and Paproth, C. and Moedas, N. and Verma, K. and Lebreton, Y. and Deal, M. and Ballot, J. and Chaplin, W. J. and Christensen-Dalsgaard, J. and Cunha, M. and Lanza, A. F. and Miglio, A. and Morel, T. and Serenelli, A. and Mosser, B. and Creevey, O. and Moya, A. and Garcia, R. A. and Nielsen, M. B. and Hatt, E.},
	month = mar,
	year = {2024},
	pages = {A78},
}

@article{lu_similarities_2022,
	title = {Similarities behind the high- and low-α disc: small intrinsic abundance scatter and migrating stars},
	volume = {512},
	issn = {0035-8711},
	shorttitle = {Similarities behind the high- and low-α disc},
	url = {https://doi.org/10.1093/mnras/stac610},
	doi = {10.1093/mnras/stac610},
	number = {2},
	urldate = {2025-05-09},
	journal = {Monthly Notices of the Royal Astronomical Society},
	author = {Lu, Yuxi (Lucy) and Ness, Melissa K and Buck, Tobias and Zinn, Joel C and Johnston, Kathryn V},
	month = may,
	year = {2022},
	pages = {2890--2910},
}

@article{casey_cannon_2016,
	title = {The {Cannon} 2: {A} data-driven model of stellar spectra for detailed chemical abundance analyses},
	shorttitle = {The {Cannon} 2},
	url = {https://ui.adsabs.harvard.edu/abs/2016arXiv160303040C/abstract},
	doi = {10.48550/arXiv.1603.03040},
	language = {en},
	urldate = {2025-05-18},
	journal = {arXiv e-prints},
	author = {Casey, Andrew R. and Hogg, David W. and Ness, Melissa and Rix, Hans-Walter and Ho, Anna Q. Y. and Gilmore, Gerry},
	month = mar,
	year = {2016},
	pages = {arXiv:1603.03040},
}

@article{lam_ages_2024,
	title = {Ages of “{Singles}” versus “{Multis}”: {Predictions} for {Dynamical} {Sculpting} over {Gyr} in the {Kepler} {Sample}},
	volume = {167},
	issn = {1538-3881},
	shorttitle = {Ages of “{Singles}” versus “{Multis}”},
	url = {https://dx.doi.org/10.3847/1538-3881/ad3804},
	doi = {10.3847/1538-3881/ad3804},
	language = {en},
	number = {6},
	urldate = {2025-05-18},
	journal = {The Astronomical Journal},
	publisher = {The American Astronomical Society},
	author = {Lam, Christopher and Ballard, Sarah},
	month = may,
	year = {2024},
	pages = {254},
}

@article{ho_label_2017,
	title = {Label {Transfer} from {APOGEE} to {LAMOST}: {Precise} {Stellar} {Parameters} for 450,000 {LAMOST} {Giants}},
	volume = {836},
	issn = {0004-637X},
	shorttitle = {Label {Transfer} from {APOGEE} to {LAMOST}},
	url = {https://dx.doi.org/10.3847/1538-4357/836/1/5},
	doi = {10.3847/1538-4357/836/1/5},
	language = {en},
	number = {1},
	urldate = {2025-05-18},
	journal = {The Astrophysical Journal},
	publisher = {The American Astronomical Society},
	author = {Ho, Anna Y. Q. and Ness, Melissa K. and Hogg, David W. and Rix, Hans-Walter and Liu, Chao and Yang, Fan and Zhang, Yong and Hou, Yonghui and Wang, Yuefei},
	month = feb,
	year = {2017},
	pages = {5},
}

@misc{behmard_data-driven_2025,
	title = {A {Data}-{Driven} {M} {Dwarf} {Model} and {Detailed} {Abundances} for \${\textbackslash}sim\$17,000 {M} {Dwarfs} in {SDSS}-{V}},
	url = {http://arxiv.org/abs/2501.14955},
	doi = {10.48550/arXiv.2501.14955},
	urldate = {2025-05-23},
	publisher = {arXiv},
	author = {Behmard, Aida and Ness, Melissa K. and Casey, Andrew R. and Angus, Ruth and Cunha, Katia and Souto, Diogo and Yuxi and Lu and Johnson, Jennifer A.},
	month = jan,
	year = {2025},
	note = {arXiv:2501.14955 [astro-ph]},
}

@article{bouma_ages_2024,
	title = {Ages of {Stars} and {Planets} in the {Kepler} {Field} {Younger} than {Four} {Billion} {Years}},
	volume = {976},
	issn = {0004-637X},
	url = {https://dx.doi.org/10.3847/1538-4357/ad855f},
	doi = {10.3847/1538-4357/ad855f},
	language = {en},
	number = {2},
	urldate = {2025-05-28},
	journal = {The Astrophysical Journal},
	publisher = {The American Astronomical Society},
	author = {Bouma, Luke G. and Hillenbrand, Lynne A. and Howard, Andrew W. and Isaacson, Howard and Masuda, Kento and Palumbo, Elsa K.},
	month = nov,
	year = {2024},
	pages = {234},
}

@article{lu_this_2024,
	title = {In {This} {Day} and {Age}: {An} {Empirical} {Gyrochronology} {Relation} for {Partially} and {Fully} {Convective} {Single} {Field} {Stars}},
	volume = {167},
	issn = {1538-3881},
	shorttitle = {In {This} {Day} and {Age}},
	url = {https://dx.doi.org/10.3847/1538-3881/ad28b9},
	doi = {10.3847/1538-3881/ad28b9},
	language = {en},
	number = {4},
	urldate = {2025-05-28},
	journal = {The Astronomical Journal},
	publisher = {The American Astronomical Society},
	author = {Lu, Yuxi(Lucy) and Angus, Ruth and Foreman-Mackey, Daniel and Hattori, Soichiro},
	month = mar,
	year = {2024},
	pages = {159},
}

@article{aerts_asteroseismology_2010,
	title = {Asteroseismology},
	url = {https://ui.adsabs.harvard.edu/abs/2010aste.book.....A/abstract},
	doi = {10.1007/978-1-4020-5803-5},
	language = {en},
	urldate = {2025-06-12},
	journal = {Asteroseismology},
	author = {Aerts, Conny and Christensen-Dalsgaard, Jørgen and Kurtz, Donald W.},
	year = {2010},
	note = {ISBN: 9781402051784},
}

@article{berger_gaia-kepler_2020,
	title = {The {Gaia}-{Kepler} {Stellar} {Properties} {Catalog}. {I}. {Homogeneous} {Fundamental} {Properties} for 186,301 {Kepler} {Stars}},
	volume = {159},
	issn = {0004-6256},
	url = {https://ui.adsabs.harvard.edu/abs/2020AJ....159..280B/abstract},
	doi = {10.3847/1538-3881/159/6/280},
	language = {en},
	number = {6},
	urldate = {2025-06-26},
	journal = {The Astronomical Journal},
	author = {Berger, Travis A. and Huber, Daniel and van Saders, Jennifer L. and Gaidos, Eric and Tayar, Jamie and Kraus, Adam L.},
	month = jun,
	year = {2020},
	pages = {280},
}

@article{chen_planets_2021,
	title = {Planets {Across} {Space} and {Time} ({PAST}). {II}. {Catalog} and {Analyses} of the {LAMOST}–{Gaia}–{Kepler} {Stellar} {Kinematic} {Properties}},
	volume = {162},
	issn = {1538-3881},
	url = {https://dx.doi.org/10.3847/1538-3881/ac0f08},
	doi = {10.3847/1538-3881/ac0f08},
	language = {en},
	number = {3},
	urldate = {2025-06-26},
	journal = {The Astronomical Journal},
	publisher = {The American Astronomical Society},
	author = {Chen, Di-Chang and Yang, Jia-Yi and Xie, Ji-Wei and Zhou, Ji-Lin and Dong, Subo and Zheng, Zheng and Zhang, Jing-Hua and Liu, Chao and Wang, Hai-Feng and Xiang, Mao-Sheng and Zong, Weikai and Huang, Yang and Luo, Ali},
	month = aug,
	year = {2021},
	pages = {100},
}

@article{aguirre_standing_2017,
	title = {Standing on the {Shoulders} of {Dwarfs}: the {Kepler} {Asteroseismic} {LEGACY} {Sample}. {II}. {Radii}, {Masses}, and {Ages}},
	volume = {835},
	issn = {0004-637X},
	shorttitle = {Standing on the {Shoulders} of {Dwarfs}},
	url = {https://dx.doi.org/10.3847/1538-4357/835/2/173},
	doi = {10.3847/1538-4357/835/2/173},
	language = {en},
	number = {2},
	urldate = {2025-07-07},
	journal = {The Astrophysical Journal},
	publisher = {The American Astronomical Society},
	author = {Aguirre, Víctor Silva and Lund, Mikkel N. and Antia, H. M. and Ball, Warrick H. and Basu, Sarbani and Christensen-Dalsgaard, Jørgen and Lebreton, Yveline and Reese, Daniel R. and Verma, Kuldeep and Casagrande, Luca and Justesen, Anders B. and Mosumgaard, Jakob R. and Chaplin, William J. and Bedding, Timothy R. and Davies, Guy R. and Handberg, Rasmus and Houdek, Günter and Huber, Daniel and Kjeldsen, Hans and Latham, David W. and White, Timothy R. and Coelho, Hugo R. and Miglio, Andrea and Rendle, Ben},
	month = jan,
	year = {2017},
	pages = {173},
}

@article{ness_spectroscopic_2016,
	title = {Spectroscopic {Determination} of {Masses} (and {Implied} {Ages}) for {Red} {Giants}},
	volume = {823},
	issn = {0004-637X},
	url = {https://ui.adsabs.harvard.edu/abs/2016ApJ...823..114N/abstract},
	doi = {10.3847/0004-637X/823/2/114},
	language = {en},
	number = {2},
	urldate = {2025-07-07},
	journal = {The Astrophysical Journal},
	author = {Ness, M. and Hogg, David W. and Rix, H.-W. and Martig, M. and Pinsonneault, Marc H. and Ho, A. Y. Q.},
	month = jun,
	year = {2016},
	pages = {114},
}

@article{montalto_all-sky_2021,
	title = {The all-sky {PLATO} input catalogue},
	volume = {653},
	issn = {0004-6361},
	url = {https://ui.adsabs.harvard.edu/abs/2021A%26A...653A..98M/abstract},
	doi = {10.1051/0004-6361/202140717},
	language = {en},
	urldate = {2025-07-10},
	journal = {Astronomy \&amp; Astrophysics, Volume 653, id.A98, {\textless}NUMPAGES{\textgreater}23{\textless}/NUMPAGES{\textgreater} pp.},
	author = {Montalto, M. and Piotto, G. and Marrese, P. M. and Nascimbeni, V. and Prisinzano, L. and Granata, V. and Marinoni, S. and Desidera, S. and Ortolani, S. and Aerts, C. and Alei, E. and Altavilla, G. and Benatti, S. and Börner, A. and Cabrera, J. and Claudi, R. and Deleuil, M. and Fabrizio, M. and Gizon, L. and Goupil, M. J. and Heras, A. M. and Magrin, D. and Malavolta, L. and Mas-Hesse, J. M. and Pagano, I. and Paproth, C. and Pertenais, M. and Pollacco, D. and Ragazzoni, R. and Ramsay, G. and Rauer, H. and Udry, S.},
	month = sep,
	year = {2021},
	pages = {A98},
}

@article{pinsonneault_apokasc_2014,
	title = {The {APOKASC} {Catalog}: {An} {Asteroseismic} and {Spectroscopic} {Joint} {Survey} of {Targets} in the {Kepler} {Fields}},
	volume = {215},
	issn = {0067-0049},
	shorttitle = {The {APOKASC} {Catalog}},
	url = {https://ui.adsabs.harvard.edu/abs/2014ApJS..215...19P/abstract},
	doi = {10.1088/0067-0049/215/2/19},
	language = {en},
	number = {2},
	urldate = {2025-07-10},
	journal = {The Astrophysical Journal Supplement Series},
	author = {Pinsonneault, Marc H. and Elsworth, Yvonne and Epstein, Courtney and Hekker, Saskia and Mészáros, Sz and Chaplin, William J. and Johnson, Jennifer A. and García, Rafael A. and Holtzman, Jon and Mathur, Savita and García Pérez, Ana and Silva Aguirre, Victor and Girardi, Léo and Basu, Sarbani and Shetrone, Matthew and Stello, Dennis and Allende Prieto, Carlos and An, Deokkeun and Beck, Paul and Beers, Timothy C. and Bizyaev, Dmitry and Bloemen, Steven and Bovy, Jo and Cunha, Katia and De Ridder, Joris and Frinchaboy, Peter M. and García-Hernández, D. A. and Gilliland, Ronald and Harding, Paul and Hearty, Fred R. and Huber, Daniel and Ivans, Inese and Kallinger, Thomas and Majewski, Steven R. and Metcalfe, Travis S. and Miglio, Andrea and Mosser, Benoit and Muna, Demitri and Nidever, David L. and Schneider, Donald P. and Serenelli, Aldo and Smith, Verne V. and Tayar, Jamie and Zamora, Olga and Zasowski, Gail},
	month = dec,
	year = {2014},
	pages = {19},
}

@article{haywood_age_2013,
	title = {The age structure of stellar populations in the solar vicinity - {Clues} of a two-phase formation history of the {Milky} {Way} disk},
	volume = {560},
	copyright = {© ESO, 2013},
	issn = {0004-6361, 1432-0746},
	url = {https://www.aanda.org/articles/aa/abs/2013/12/aa21397-13/aa21397-13.html},
	doi = {10.1051/0004-6361/201321397},
	language = {en},
	urldate = {2025-07-16},
	journal = {Astronomy \& Astrophysics},
	publisher = {EDP Sciences},
	author = {Haywood, Misha and Matteo, Paola Di and Lehnert, Matthew D. and Katz, David and Gómez, Ana},
	month = dec,
	year = {2013},
	pages = {A109},
}

@article{meszaros_sdss-v_2025,
	title = {{SDSS}-{V} {Milky} {Way} {Mapper} ({MWM}): {ASPCAP} {Stellar} {Parameters} and {Abundances} in {SDSS}-{V} {Data} {Release} 19},
	volume = {170},
	issn = {0004-6256},
	shorttitle = {{SDSS}-{V} {Milky} {Way} {Mapper} ({MWM})},
	url = {https://ui.adsabs.harvard.edu/abs/2025AJ....170...96M/abstract},
	doi = {10.3847/1538-3881/ade4b9},
	language = {en},
	number = {2},
	urldate = {2025-08-19},
	journal = {The Astronomical Journal},
	author = {Mészáros, Szabolcs and Jofré, Paula and Johnson, Jennifer A. and Bird, Jonathan C. and Bovy, Jo and Casey, Andrew R. and Chanamé, Julio and Cunha, Katia and De Lee, Nathan and Frinchaboy, Peter and Guiglion, Guillaume and Hegedűs, Viola and Ji, Alex P. and Kollmeier, Juna A. and Ness, Melissa K. and Otto, Jonah and Pinsonneault, Marc H. and Roman-Lopes, Alexandre and Saydjari, Andrew and Sinha, Amaya and Song, Ying-Yi and Stringfellow, Guy S. and Stassun, Keivan G. and Tayar, Jamie and Tkachenko, Andrew and Valentini, Marica and Way, Zachary and Weingrill, Jörg},
	month = aug,
	year = {2025},
	pages = {96},
}

@article{zhang_most_2021,
	title = {Most "young" \$α\$-rich stars have high masses but are actually old},
	volume = {922},
	issn = {0004-637X, 1538-4357},
	url = {http://arxiv.org/abs/2109.00746},
	doi = {10.3847/1538-4357/ac22a5},
	number = {2},
	urldate = {2025-08-21},
	journal = {The Astrophysical Journal},
	author = {Zhang, Meng and Xiang, Mao-Sheng and Zhang, Hua-Wei and Ting, Yuan-Sen and Rix, Hans-Walter and Wu, Ya-qian and Huang, Yang and Sun, Wei-Xiang and Tian, Zhi-Jia and Wang, Chun and Liu, Xiao-Wei},
	month = dec,
	year = {2021},
	note = {arXiv:2109.00746 [astro-ph]},
	pages = {145},
}

@misc{lam_late-time_2025,
	title = {A {Late}-{Time} {Rise} in {Planet} {Occurrence} {Reproduces} the {Galactic} {Height} {Trend} in {Planet} {Occurrence}},
	url = {http://arxiv.org/abs/2507.21250},
	doi = {10.48550/arXiv.2507.21250},
	urldate = {2025-09-01},
	publisher = {arXiv},
	author = {Lam, Christopher and Ballard, Sarah and Sagear, Sheila and Daniel, Kathryne J.},
	month = jul,
	year = {2025},
	note = {arXiv:2507.21250 [astro-ph]},
}

@article{sayeed_homogeneous_2025,
	title = {A {Homogeneous} {Catalog} of {Oscillating} {Solar}-type {Stars} {Observed} by the {Kepler} {Mission} and a {New} {Amplitude} {Scaling} {Relation} {Including} {Chromospheric} {Activity}},
	volume = {170},
	issn = {1538-3881},
	url = {https://doi.org/10.3847/1538-3881/adf648},
	doi = {10.3847/1538-3881/adf648},
	language = {en},
	number = {4},
	urldate = {2025-09-26},
	journal = {The Astronomical Journal},
	publisher = {The American Astronomical Society},
	author = {Sayeed, Maryum and Huber, Daniel and Chontos, Ashley and Li, Yaguang},
	month = sep,
	year = {2025},
	pages = {212},
}

@article{boettner_exoplanets_2024,
	title = {Exoplanets across galactic stellar populations with {PLATO}: {Estimating} exoplanet yields around {FGK} stars for the thin disk, thick disk, and stellar halo},
	volume = {692},
	copyright = {https://creativecommons.org/licenses/by/4.0},
	issn = {0004-6361, 1432-0746},
	shorttitle = {Exoplanets across galactic stellar populations with {PLATO}},
	url = {https://www.aanda.org/10.1051/0004-6361/202451537},
	doi = {10.1051/0004-6361/202451537},
	language = {en},
	urldate = {2025-11-13},
	journal = {Astronomy \& Astrophysics},
	author = {Boettner, C. and Viswanathan, A. and Dayal, P.},
	month = dec,
	year = {2024},
	pages = {A150},
}

@article{chaplin_asteroseismology_2013,
	title = {Asteroseismology of {Solar}-{Type} and {Red}-{Giant} {Stars}},
	volume = {51},
	issn = {0066-4146, 1545-4282},
	url = {https://www.annualreviews.org/content/journals/10.1146/annurev-astro-082812-140938},
	doi = {10.1146/annurev-astro-082812-140938},
	language = {en},
	number = {Volume 51, 2013},
	urldate = {2026-02-11},
	journal = {Annual Review of Astronomy and Astrophysics},
	publisher = {Annual Reviews},
	author = {Chaplin, William J. and Miglio, Andrea},
	month = aug,
	year = {2013},
	pages = {353--392},
}

@article{soderblom_ages_2010,
	title = {The {Ages} of {Stars}},
	volume = {48},
	issn = {0066-4146, 1545-4282},
	url = {https://www.annualreviews.org/content/journals/10.1146/annurev-astro-081309-130806},
	doi = {10.1146/annurev-astro-081309-130806},
	language = {en},
	number = {Volume 48, 2010},
	urldate = {2026-02-11},
	journal = {Annual Review of Astronomy and Astrophysics},
	publisher = {Annual Reviews},
	author = {Soderblom, David R.},
	month = sep,
	year = {2010},
	pages = {581--629},
}

@article{lebreton_stellar_2008,
	title = {Stellar ages from asteroseismology},
	volume = {4},
	copyright = {https://www.cambridge.org/core/terms},
	issn = {1743-9213, 1743-9221},
	url = {https://www.cambridge.org/core/product/identifier/S1743921309032074/type/journal_article},
	doi = {10.1017/S1743921309032074},
	language = {en},
	number = {S258},
	urldate = {2026-02-11},
	journal = {Proceedings of the International Astronomical Union},
	author = {Lebreton, Yveline and Montalbán, Josefina},
	month = oct,
	year = {2008},
	pages = {419--430},
}

@misc{rauer_plato_2024,
	title = {The {PLATO} {Mission}},
	url = {http://arxiv.org/abs/2406.05447},
	doi = {10.48550/arXiv.2406.05447},
	urldate = {2026-02-11},
	publisher = {arXiv},
	author = {Rauer, Heike and Aerts, Conny and Cabrera, Juan and Deleuil, Magali and Erikson, Anders and Gizon, Laurent and Goupil, Mariejo and Heras, Ana and Lorenzo-Alvarez, Jose and Marliani, Filippo and Martin-Garcia, César and Mas-Hesse, J. Miguel and O'Rourke, Laurence and Osborn, Hugh and Pagano, Isabella and Piotto, Giampaolo and Pollacco, Don and Ragazzoni, Roberto and Ramsay, Gavin and Udry, Stéphane and Appourchaux, Thierry and Benz, Willy and Brandeker, Alexis and Güdel, Manuel and Janot-Pacheco, Eduardo and Kabath, Petr and Kjeldsen, Hans and Min, Michiel and Santos, Nuno and Smith, Alan and Suarez, Juan-Carlos and Werner, Stephanie C. and Aboudan, Alessio and Abreu, Manuel and a, Lorena Acu and Adams, Moritz and Adibekyan, Vardan and Affer, Laura and Agneray, François and Agnor, Craig and Børsen-Koch, Victor Aguirre and Ahmed, Saad and Aigrain, Suzanne and Al-Bahlawan, Ashraf and Gil, M. de los Angeles Alcacera and Alei, Eleonora and Alencar, Silvia and Alexander, Richard and Alfonso-Garzón, Julia and Alibert, Yann and Prieto, Carlos Allende and Almeida, Leonardo and Sobrino, Roi Alonso and Altavilla, Giuseppe and Althaus, Christian and Trujillo, Luis Alonso Alvarez and Amarsi, Anish and Eiff, Matthias Ammler-von and Amôres, Eduardo and Andrade, Laerte and Antoniadis-Karnavas, Alexandros and António, Carlos and Moral, Beatriz Aparicio del and Appolloni, Matteo and Arena, Claudio and Armstrong, David and Aliaga, Jose Aroca and Asplund, Martin and Audenaert, Jeroen and Auricchio, Natalia and Avelino, Pedro and Baeke, Ann and Baillié, Kevin and Balado, Ana and Balagueró, Pau Ballber and Balestra, Andrea and Ball, Warrick and Ballans, Herve and Ballot, Jerome and Barban, Caroline and Barbary, Gaële and Barbieri, Mauro and Forteza, Sebasti Barceló and Barker, Adrian and Barklem, Paul and Barnes, Sydney and Navascues, David Barrado and Barragan, Oscar and Baruteau, Clément and Basu, Sarbani and Baudin, Frederic and Baumeister, Philipp and Bayliss, Daniel and Bazot, Michael and Beck, Paul G. and Bedding, Tim and Belkacem, Kevin and Bellinger, Earl and Benatti, Serena and Benomar, Othman and Bérard, Diane and Bergemann, Maria and Bergomi, Maria and Bernardo, Pierre and Biazzo, Katia and Bignamini, Andrea and Bigot, Lionel and Billot, Nicolas and Binet, Martin and Biondi, David and Biondi, Federico and Birch, Aaron C. and Bitsch, Bertram and Ceballos, Paz Victoria Bluhm and Bódi, Attila and Bognár, Zsófia and Boisse, Isabelle and Bolmont, Emeline and Bonanno, Alfio and Bonavita, Mariangela and Bonfanti, Andrea and Bonfils, Xavier and Bonito, Rosaria and Bonomo, Aldo Stefano and Börner, Anko and Saikia, Sudeshna Boro and Martín, Elisa Borreguero and Borsa, Francesco and Borsato, Luca and Bossini, Diego and Bouchy, Francois and Boué, Gwenaël and Boufleur, Rodrigo and Boumier, Patrick and Bourrier, Vincent and Bowman, Dominic M. and Bozzo, Enrico and Bradley, Louisa and Bray, John and Bressan, Alessandro and Breton, Sylvain and Brienza, Daniele and Brito, Ana and Brogi, Matteo and Brown, Beverly and Brown, David J. A. and Brun, Allan Sacha and Bruno, Giovanni and Bruns, Michael and Buchhave, Lars A. and Bugnet, Lisa and Buldgen, Gaël and Burgess, Patrick and Busatta, Andrea and Busso, Giorgia and Buzasi, Derek and Caballero, José A. and Cabral, Alexandre and Gomez, Juan-Francisco Cabrero and Calderone, Flavia and Cameron, Robert and Cameron, Andrew and Campante, Tiago and Gestal, Néstor Campos and Martins, Bruno Leonardo Canto and Cara, Christophe and Carone, Ludmila and Carrasco, Josep Manel and Casagrande, Luca and Casewell, Sarah L. and Cassisi, Santi and Castellani, Marco and Castro, Matthieu and Catala, Claude and Fernández, Irene Catalán and Catelan, Márcio and Cegla, Heather and Cerruti, Chiara and Cessa, Virginie and Chadid, Merieme and Chaplin, William and Charpinet, Stephane and Chiappini, Cristina and Chiarucci, Simone and Chiavassa, Andrea and Chinellato, Simonetta and Chirulli, Giovanni and Christensen-Dalsgaard, Jørgen and Church, Ross and Claret, Antonio and Clarke, Cathie and Claudi, Riccardo and Clermont, Lionel and Coelho, Hugo and Coelho, Joao and Cogato, Fabrizio and Colomé, Josep and Condamin, Mathieu and García, Fernando Conde and Conseil, Simon and Corbard, Thierry and Correia, Alexandre C. M. and Corsaro, Enrico and Cosentino, Rosario and Costes, Jean and Cottinelli, Andrea and Covone, Giovanni and Creevey, Orlagh L. and Crida, Aurelien and Csizmadia, Szilard and Cunha, Margarida and Curry, Patrick and Costa, Jefferson da and Silva, Francys da and Dalal, Shweta and Damasso, Mario and Damiani, Cilia and Damiani, Francesco and Chagas, Maria Liduina das and Davies, Melvyn and Davies, Guy and Davies, Ben and Davison, Gary and Almeida, Leandro de and Angeli, Francesca de and Barros, Susana Cristina Cabral de and Leão, Izan de Castro and Freitas, Daniel Brito de and Freitas, Marcia Cristina de and Martino, Domitilla De and Medeiros, José Renan de and Paula, Luiz Alberto de and Gómez, Álvaro de Pedraza and Plaa, Jelle de and Ridder, Joris De and Deal, Morgan and Decin, Leen and Deeg, Hans and Innocenti, Scilla Degl and Deheuvels, Sebastien and Burgo, Carlos del and Sordo, Fabio Del and Delgado-Mena, Elisa and Demangeon, Olivier and Denk, Tilmann and Derekas, Aliz and Desert, Jean-Michel and Desidera, Silvano and Dexet, Marc and Criscienzo, Marcella Di and Giorgio, Anna Maria Di and Mauro, Maria Pia Di and Rial, Federico Jose Diaz and Díaz-García, José-Javier and Dima, Marco and Dinuzzi, Giacomo and Dionatos, Odysseas and Distefano, Elisa and Nascimento, Jose-Dias do and Domingo, Albert and D'Orazi, Valentina and Dorn, Caroline and Doyle, Lauren and Duarte, Elena and Ducellier, Florent and Dumaye, Luc and Dumusque, Xavier and Dupret, Marc-Antoine and Eggenberger, Patrick and Ehrenreich, David and Eigmüller, Philipp and Eising, Johannes and Emilio, Marcelo and Eriksson, Kjell and Ermocida, Marco and Giribaldi, Riano Isidoro Escate and Eschen, Yoshi and ez, Lucía Espinosa Yá and Estrela, In s and Evans, Dafydd Wyn and Fabbian, Damian and Fabrizio, Michele and Faria, João Pedro and Farina, Maria and Farinato, Jacopo and Feliz, Dax and Feltzing, Sofia and Fenouillet, Thomas and Fernández, Miguel and Ferrari, Lorenza and Ferraz-Mello, Sylvio and Fialho, Fabio and Fienga, Agnes and Figueira, Pedro and Fiori, Laura and Flaccomio, Ettore and Focardi, Mauro and Foley, Steve and Fontignie, Jean and Ford, Dominic and Fornazier, Karin and Forveille, Thierry and Fossati, Luca and Franca, Rodrigo de Marca and Silva, Lucas Franco da and Frasca, Antonio and Fridlund, Malcolm and Furlan, Marco and Gabler, Sarah-Maria and Gaido, Marco and Gallagher, Andrew and Sempere, Paloma I. Gallego and Galli, Emanuele and García, Rafael A. and Hernández, Antonio García and Munoz, Antonio Garcia and García-Vázquez, Hugo and Haba, Rafael Garrido and Gaulme, Patrick and Gauthier, Nicolas and Gehan, Charlotte and Gent, Matthew and Georgieva, Iskra and Ghigo, Mauro and Giana, Edoardo and Gill, Samuel and Girardi, Leo and Winter, Silvia Giuliatti and Giusi, Giovanni and Silva, João Gomes da and Zazo, Luis Jorge Gómez and Gomez-Lopez, Juan Manuel and Hernández, Jonay Isai González and Murillo, Kevin Gonzalez and Melchor, Alejandro Gonzalo and Gorius, Nicolas and Gouel, Pierre-Vincent and Goulty, Duncan and Granata, Valentina and Grenfell, John Lee and bach, Denis Grie and Grolleau, Emmanuel and Grouffal, Salomé and Grziwa, Sascha and Guarcello, Mario Giuseppe and Gueguen, Lo c and Guenther, Eike Wolf and Guilhem, Terrasa and Guillerot, Lucas and Guillot, Tristan and Guiot, Pierre and Guterman, Pascal and Gutiérrez, Antonio and Gutiérrez-Canales, Fernando and Hagelberg, Janis and Haldemann, Jonas and Hall, Cassandra and Handberg, Rasmus and Harrison, Ian and Harrison, Diana L. and Hasiba, Johann and Haswell, Carole A. and Hatalova, Petra and Hatzes, Artie and Haywood, Raphaelle and Hébrard, Guillaume and Heckes, Frank and Heiter, Ulrike and Hekker, Saskia and Heller, René and Helling, Christiane and Helminiak, Krzysztof and Hemsley, Simon and Heng, Kevin and Herbst, Konstantin and Hermans, Aline and Hermes, J. J. and Torres, Nadia Hidalgo and Hinkel, Natalie and Hobbs, David and Hodgkin, Simon and Hofmann, Karl and Hojjatpanah, Saeed and Houdek, Günter and Huber, Daniel and Huesler, Joseph and Hui-Bon-Hoa, Alain and Huygen, Rik and Huynh, Duc-Dat and Iro, Nicolas and Irwin, Jonathan and Irwin, Mike and Izidoro, André and Jacquinod, Sophie and Jannsen, Nicholas Emborg and Janson, Markus and Jeszenszky, Harald and Jiang, Chen and Mancebo, Antonio José Jimenez and Jofre, Paula and Johansen, Anders and Johnston, Cole and Jones, Geraint and Kallinger, Thomas and Kálmán, Szilárd and Kanitz, Thomas and Karjalainen, Marie and Karjalainen, Raine and Karoff, Christoffer and Kawaler, Steven and Kawata, Daisuke and Keereman, Arnoud and Keiderling, David and Kennedy, Tom and Kenworthy, Matthew and Kerschbaum, Franz and Kidger, Mark and Kiefer, Flavien and Kintziger, Christian and Kislyakova, Kristina and Kiss, László and Klagyivik, Peter and Klahr, Hubert and Klevas, Jonas and Kochukhov, Oleg and Köhler, Ulrich and Kolb, Ulrich and Koncz, Alexander and Korth, Judith and Kostogryz, Nadiia and Kovács, Gábor and Kovács, József and Kozhura, Oleg and Krivova, Natalie and Kucinskas, Arunas and Kuhlemann, Ilyas and Kupka, Friedrich and Laauwen, Wouter and Labiano, Alvaro and Lagarde, Nadege and Laget, Philippe and Laky, Gunter and Lam, Kristine Wai Fun and Lambrechts, Michiel and Lammer, Helmut and Lanza, Antonino Francesco and Lanzafame, Alessandro and Martiz, Mariel Lares and Laskar, Jacques and Latter, Henrik and Lavanant, Tony and Lawrenson, Alastair and Lazzoni, Cecilia and Lebre, Agnes and Lebreton, Yveline and Etangs, Alain Lecavelier des and Lee, Katherine and Leinhardt, Zoe and Leleu, Adrien and Lendl, Monika and Leto, Giuseppe and Levillain, Yves and Libert, Anne-Sophie and Lichtenberg, Tim and Ligi, Roxanne and Lignieres, Francois and Lillo-Box, Jorge and Linsky, Jeffrey and Liu, John Scige and Loidolt, Dominik and Longval, Yuying and Lopes, Ilídio and Lorenzani, Andrea and Ludwig, Hans-Guenter and Lund, Mikkel and Lundkvist, Mia Sloth and Luri, Xavier and Maceroni, Carla and Madden, Sean and Madhusudhan, Nikku and Maggio, Antonio and Magliano, Christian and Magrin, Demetrio and Mahy, Laurent and Maibaum, Olaf and Malac-Allain, LeeRoy and Malapert, Jean-Christophe and Malavolta, Luca and Maldonado, Jesus and Mamonova, Elena and Manchon, Louis and Manjón, Andres and Mann, Andrew and Mantovan, Giacomo and Marafatto, Luca and Marconi, Marcella and Mardling, Rosemary and Marigo, Paola and Marinoni, Silvia and Marques, rico and Marques, Joao Pedro and Marrese, Paola Maria and Marshall, Douglas and Perales, Silvia Martínez and Mary, David and Marzari, Francesco and Masana, Eduard and Mascher, Andrina and Mathis, Stéphane and Mathur, Savita and Vodopivec, Iris Martín and Figueiredo, Ana Carolina Mattiuci and Maxted, Pierre F. L. and Mazeh, Tsevi and Mazevet, Stephane and Mazzei, Francesco and McCormac, James and McMillan, Paul and Menou, Lucas and Merle, Thibault and Meru, Farzana and Mesa, Dino and Messina, Sergio and Mészáros, Szabolcs and Meunier, Nadége and Meunier, Jean-Charles and Micela, Giuseppina and Michaelis, Harald and Michel, Eric and Michielsen, Mathias and Michtchenko, Tatiana and Miglio, Andrea and Miguel, Yamila and Milligan, David and Mirouh, Giovanni and Mitchell, Morgan and Moedas, Nuno and Molendini, Francesca and Molnár, László and Mombarg, Joey and Montalban, Josefina and Montalto, Marco and Monteiro, Mário J. P. F. G. and Sánchez, Francisco Montoro and Morales, Juan Carlos and Morales-Calderon, Maria and Morbidelli, Alessandro and Mordasini, Christoph and Moreau, Chrystel and Morel, Thierry and Morello, Guiseppe and Morin, Julien and Mortier, Annelies and Mosser, Beno t and Mourard, Denis and Mousis, Olivier and Moutou, Claire and Mowlavi, Nami and Moya, Andrés and Muehlmann, Prisca and Muirhead, Philip and Munari, Matteo and Musella, Ilaria and Mustill, Alexander James and Nardetto, Nicolas and Nardiello, Domenico and Narita, Norio and Nascimbeni, Valerio and Nash, Anna and Neiner, Coralie and Nelson, Richard P. and Nettelmann, Nadine and Nicolini, Gianalfredo and Nielsen, Martin and Niemi, Sami-Matias and Noack, Lena and Noels-Grotsch, Arlette and Noll, Anthony and Norazman, Azib and Norton, Andrew J. and Nsamba, Benard and Ofir, Aviv and Ogilvie, Gordon and Olander, Terese and Olivetto, Christian and Olofsson, Göran and Ong, Joel and Ortolani, Sergio and Oshagh, Mahmoudreza and Ottacher, Harald and Ottensamer, Roland and Ouazzani, Rhita-Maria and Paardekooper, Sijme-Jan and Pace, Emanuele and Pajas, Miriam and Palacios, Ana and Palandri, Gaelle and Palle, Enric and Paproth, Carsten and Parro, Vanderlei and Parviainen, Hannu and Granado, Javier Pascual and Passegger, Vera Maria and Pastor-Morales, Carmen and Pätzold, Martin and Pedersen, May Gade and Hidalgo, David Pena and Pepe, Francesco and Pereira, Filipe and Persson, Carina M. and Pertenais, Martin and Peter, Gisbert and Petit, Antoine C. and Petit, Pascal and Pezzuto, Stefania and Pichierri, Gabriele and Pietrinferni, Adriano and Pinheiro, Fernando and Pinsonneault, Marc and Plachy, Emese and Plasson, Philippe and Plez, Bertrand and Poppenhaeger, Katja and Poretti, Ennio and Portaluri, Elisa and Portell, Jordi and Mello, Gustavo Frederico Porto de and Poyatos, Julien and Pozuelos, Francisco J. and Moroni, Pier Giorgio Prada and Pricopi, Dumitru and Prisinzano, Loredana and Quade, Matthias and Quirrenbach, Andreas and Reina, Julio Arturo Rabanal and Soares, Maria Cristina Rabello and Raimondo, Gabriella and Rainer, Monica and Rodón, Jose Ramón and Ramón-Ballesta, Alejandro and Zapata, Gonzalo Ramos and Rätz, Stefanie and Rauterberg, Christoph and Redman, Bob and Redmer, Ronald and Reese, Daniel and Regibo, Sara and Reiners, Ansgar and Reinhold, Timo and Renie, Christian and Ribas, Ignasi and Ribeiro, Sergio and Ricciardi, Thiago Pereira and Rice, Ken and Richard, Olivier and Riello, Marco and Rieutord, Michel and Ripepi, Vincenzo and Rixon, Guy and Rockstein, Steve and Ortiz, José Ramón Rodón and Rodríguez, María Teresa Rodrigo and Amor, Alberto Rodríguez and Díaz, Luisa Fernanda Rodríguez and Garcia, Juan Pablo Rodriguez and Rodriguez-Gomez, Julio and Roehlly, Yannick and Roig, Fernando and Rojas-Ayala, Bárbara and Rolf, Tobias and Rørsted, Jakob Lysgaard and Rosado, Hugo and Rosotti, Giovanni and Roth, Olivier and Roth, Markus and Rousseau, Alex and Roxburgh, Ian and Roy, Fabrice and Royer, Pierre and Ruane, Kirk and Mastropasqua, Sergio Rufini and Galarreta, Claudia Ruiz de and Russi, Andrea and Saar, Steven and Saillenfest, Melaine and Salaris, Maurizio and Salmon, Sebastien and Saltas, Ippocratis and Samadi, Réza and Samadi, Aunia and Samra, Dominic and Silva, Tiago Sanches da and Carrasco, Miguel Andrés Sánchez and Santerne, Alexandre and Pé, Amaia Santiago and Santoli, Francesco and Santos, ngela R. G. and Mesa, Rosario Sanz and Sarro, Luis Manuel and Scandariato, Gaetano and Schäfer, Martin and Schlafly, Edward and Schmider, François-Xavier and Schneider, Jean and Schou, Jesper and Schunker, Hannah and Schwarzkopf, Gabriel Jörg and Serenelli, Aldo and Seynaeve, Dries and Shan, Yutong and Shapiro, Alexander and Shipman, Russel and Sicilia, Daniela and sanmartin, Maria Angeles Sierra and Sigot, Axelle and Silliman, Kyle and Silvotti, Roberto and Simon, Attila E. and Napoli, Ricardo Simoyama and Skarka, Marek and Smalley, Barry and Smiljanic, Rodolfo and Smit, Samuel and Smith, Alexis and Smith, Leigh and Snellen, Ignas and Sódor, Ádám and Sohl, Frank and Solanki, Sami K. and Sortino, Francesca and Sousa, Sérgio and Southworth, John and Souto, Diogo and Sozzetti, Alessandro and Stamatellos, Dimitris and Stassun, Keivan and Steller, Manfred and Stello, Dennis and Stelzer, Beate and Stiebeler, Ulrike and Stokholm, Amalie and Storelvmo, Trude and Strassmeier, Klaus and Strøm, Paul Anthony and Strugarek, Antoine and Sulis, Sophia and vanda, Michal and Szabados, László and Szabó, Róbert and Szabó, Gyula M. and Szuszkiewicz, Ewa and Talens, Geert Jan and Teti, Daniele and Theisen, Tom and Thévenin, Frédéric and Thoul, Anne and Tiphene, Didier and Titz-Weider, Ruth and Tkachenko, Andrew and Tomecki, Daniel and Tonfat, Jorge and Tosi, Nicola and Trampedach, Regner and Traven, Gregor and Triaud, Amaury and Trønnes, Reidar and Tsantaki, Maria and Tschentscher, Matthias and Turin, Arnaud and Tvaruzka, Adam and Ulmer, Bernd and Ulmer-Moll, Solène and Ulusoy, Ceren and Umbriaco, Gabriele and Valencia, Diana and Valentini, Marica and Valio, Adriana and Guijarro, Ángel Luis Valverde and Eylen, Vincent Van and Grootel, Valerie Van and Kempen, Tim A. van and Reeth, Timothy Van and Zelst, Iris Van and Vandenbussche, Bart and Vasiliou, Konstantinos and Vasilyev, Valeriy and Mascarenhas, David Vaz de and Vazan, Allona and Nunez, Marina Vela and Velloso, Eduardo Nunes and Ventura, Rita and Ventura, Paolo and Venturini, Julia and Trallero, Isabel Vera and Veras, Dimitri and Verdugo, Eva and Verma, Kuldeep and Vibert, Didier and Martinez, Tobias Vicanek and Vida, Krisztián and Vigan, Arthur and Villacorta, Antonio and Villaver, Eva and Aparicio, Marcos Villaverde and Viotto, Valentina and Vorobyov, Eduard and Vorontsov, Sergey and Wagner, Frank W. and Walloschek, Thomas and Walton, Nicholas and Walton, Dave and Wang, Haiyang and Waters, Rens and Watson, Christopher and Wedemeyer, Sven and Weeks, Angharad and Weingrill, Jörg and Weiss, Annita and Wendler, Belinda and West, Richard and Westerdorff, Karsten and Westphal, Pierre-Amaury and Wheatley, Peter and White, Tim and Whittaker, Amadou and Wickhusen, Kai and Wilson, Thomas and Windsor, James and Winter, Othon and Winther, Mark Lykke and Winton, Alistair and Witteck, Ulrike and Witzke, Veronika and Woitke, Peter and Wolter, David and Wuchterl, Günther and Wyatt, Mark and Yang, Dan and Yu, Jie and Sanchez, Ricardo Zanmar and Osorio, María Rosa Zapatero and Zechmeister, Mathias and Zhou, Yixiao and Ziemke, Claas and Zwintz, Konstanze},
	month = nov,
	year = {2024},
	note = {arXiv:2406.05447 [astro-ph]},
}

@article{maas_exploring_2025,
	title = {Exploring the {History} of {Stellar} {Mergers} with {Chemistry}: {Examining} the {Origins} of {Massive} α-enriched {Stars} {Using} {Carbon} {Isotope} {Ratios}},
	volume = {994},
	issn = {0004-637X},
	shorttitle = {Exploring the {History} of {Stellar} {Mergers} with {Chemistry}},
	url = {https://doi.org/10.3847/1538-4357/ae0a2d},
	doi = {10.3847/1538-4357/ae0a2d},
	language = {en},
	number = {1},
	urldate = {2026-02-14},
	journal = {The Astrophysical Journal},
	publisher = {The American Astronomical Society},
	author = {Maas, Zachary G. and Hawkins, Keith and Gerber, Jeffrey M. and Hackshaw, Zoe and Manea, Catherine},
	month = nov,
	year = {2025},
	pages = {16},
}

@article{angus_toward_2019,
	title = {Toward {Precise} {Stellar} {Ages}: {Combining} {Isochrone} {Fitting} with {Empirical} {Gyrochronology}},
	volume = {158},
	issn = {1538-3881},
	shorttitle = {Toward {Precise} {Stellar} {Ages}},
	url = {https://doi.org/10.3847/1538-3881/ab3c53},
	doi = {10.3847/1538-3881/ab3c53},
	language = {en},
	number = {5},
	urldate = {2026-03-21},
	journal = {The Astronomical Journal},
	publisher = {The American Astronomical Society},
	author = {Angus, Ruth and Morton, Timothy D. and Foreman-Mackey, Daniel and van Saders, Jennifer and Curtis, Jason and Kane, Stephen R. and Bedell, Megan and Kiman, Rocio and Hogg, David W. and Brewer, John},
	month = oct,
	year = {2019},
	pages = {173},
}

@article{sahlholdt_benchmark_2019,
	title = {Benchmark ages for the {Gaia} benchmark stars},
	volume = {482},
	issn = {0035-8711},
	url = {https://doi.org/10.1093/mnras/sty2732},
	doi = {10.1093/mnras/sty2732},
	number = {1},
	urldate = {2026-03-23},
	journal = {Monthly Notices of the Royal Astronomical Society},
	author = {Sahlholdt, Christian L and Feltzing, Sofia and Lindegren, Lennart and Church, Ross P},
	month = jan,
	year = {2019},
	pages = {895--920},
}

@article{sagear_zoomies_2024,
	title = {zoomies: {A} {Tool} to {Infer} {Stellar} {Age} from {Vertical} {Action} in {Gaia} {Data}},
	volume = {977},
	issn = {0004-637X},
	shorttitle = {zoomies},
	url = {https://doi.org/10.3847/1538-4357/ad8b26},
	doi = {10.3847/1538-4357/ad8b26},
	language = {en},
	number = {1},
	urldate = {2026-03-24},
	journal = {The Astrophysical Journal},
	publisher = {The American Astronomical Society},
	author = {Sagear, Sheila and Price-Whelan, Adrian M. and Ballard, Sarah and Lu, Yuxi (Lucy) and Angus, Ruth and Hogg, David W.},
	month = dec,
	year = {2024},
	pages = {49},
}

@article{schmidt_resonant_2024,
	title = {Resonant and {Ultra}-short-period {Planet} {Systems} {Are} at {Opposite} {Ends} of the {Exoplanet} {Age} {Distribution}},
	volume = {168},
	issn = {1538-3881},
	url = {https://doi.org/10.3847/1538-3881/ad5d76},
	doi = {10.3847/1538-3881/ad5d76},
	language = {en},
	number = {3},
	urldate = {2026-03-24},
	journal = {The Astronomical Journal},
	publisher = {The American Astronomical Society},
	author = {Schmidt, Stephen P. and Schlaufman, Kevin C. and Hamer, Jacob H.},
	month = aug,
	year = {2024},
	pages = {109},
}

@article{vallenari_gaia_2023,
	title = {Gaia {Data} {Release} 3 - {Summary} of the content and survey properties},
	volume = {674},
	copyright = {© The Authors 2023},
	issn = {0004-6361, 1432-0746},
	url = {https://www.aanda.org/articles/aa/abs/2023/06/aa43940-22/aa43940-22.html},
	doi = {10.1051/0004-6361/202243940},
	language = {en},
	urldate = {2026-03-24},
	journal = {Astronomy \& Astrophysics},
	publisher = {EDP Sciences},
	author = {Vallenari, A. and Brown, A. G. A. and Prusti, T. and Bruijne, J. H. J. de and Arenou, F. and Babusiaux, C. and Biermann, M. and Creevey, O. L. and Ducourant, C. and Evans, D. W. and Eyer, L. and Guerra, R. and Hutton, A. and Jordi, C. and Klioner, S. A. and Lammers, U. L. and Lindegren, L. and Luri, X. and Mignard, F. and Panem, C. and Pourbaix, D. and Randich, S. and Sartoretti, P. and Soubiran, C. and Tanga, P. and Walton, N. A. and Bailer-Jones, C. a. L. and Bastian, U. and Drimmel, R. and Jansen, F. and Katz, D. and Lattanzi, M. G. and Leeuwen, F. van and Bakker, J. and Cacciari, C. and Castañeda, J. and Angeli, F. De and Fabricius, C. and Fouesneau, M. and Frémat, Y. and Galluccio, L. and Guerrier, A. and Heiter, U. and Masana, E. and Messineo, R. and Mowlavi, N. and Nicolas, C. and Nienartowicz, K. and Pailler, F. and Panuzzo, P. and Riclet, F. and Roux, W. and Seabroke, G. M. and Sordo, R. and Thévenin, F. and Gracia-Abril, G. and Portell, J. and Teyssier, D. and Altmann, M. and Andrae, R. and Audard, M. and Bellas-Velidis, I. and Benson, K. and Berthier, J. and Blomme, R. and Burgess, P. W. and Busonero, D. and Busso, G. and Cánovas, H. and Carry, B. and Cellino, A. and Cheek, N. and Clementini, G. and Damerdji, Y. and Davidson, M. and Teodoro, P. de and Campos, M. Nuñez and Delchambre, L. and Dell’Oro, A. and Esquej, P. and Fernández-Hernández, J. and Fraile, E. and Garabato, D. and García-Lario, P. and Gosset, E. and Haigron, R. and Halbwachs, J.-L. and Hambly, N. C. and Harrison, D. L. and Hernández, J. and Hestroffer, D. and Hodgkin, S. T. and Holl, B. and Janßen, K. and Fombelle, G. Jevardat de and Jordan, S. and Krone-Martins, A. and Lanzafame, A. C. and Löffler, W. and Marchal, O. and Marrese, P. M. and Moitinho, A. and Muinonen, K. and Osborne, P. and Pancino, E. and Pauwels, T. and Recio-Blanco, A. and Reylé, C. and Riello, M. and Rimoldini, L. and Roegiers, T. and Rybizki, J. and Sarro, L. M. and Siopis, C. and Smith, M. and Sozzetti, A. and Utrilla, E. and Leeuwen, M. van and Abbas, U. and Ábrahám, P. and Aramburu, A. Abreu and Aerts, C. and Aguado, J. J. and Ajaj, M. and Aldea-Montero, F. and Altavilla, G. and Álvarez, M. A. and Alves, J. and Anders, F. and Anderson, R. I. and Varela, E. Anglada and Antoja, T. and Baines, D. and Baker, S. G. and Balaguer-Núñez, L. and Balbinot, E. and Balog, Z. and Barache, C. and Barbato, D. and Barros, M. and Barstow, M. A. and Bartolomé, S. and Bassilana, J.-L. and Bauchet, N. and Becciani, U. and Bellazzini, M. and Berihuete, A. and Bernet, M. and Bertone, S. and Bianchi, L. and Binnenfeld, A. and Blanco-Cuaresma, S. and Blazere, A. and Boch, T. and Bombrun, A. and Bossini, D. and Bouquillon, S. and Bragaglia, A. and Bramante, L. and Breedt, E. and Bressan, A. and Brouillet, N. and Brugaletta, E. and Bucciarelli, B. and Burlacu, A. and Butkevich, A. G. and Buzzi, R. and Caffau, E. and Cancelliere, R. and Cantat-Gaudin, T. and Carballo, R. and Carlucci, T. and Carnerero, M. I. and Carrasco, J. M. and Casamiquela, L. and Castellani, M. and Castro-Ginard, A. and Chaoul, L. and Charlot, P. and Chemin, L. and Chiaramida, V. and Chiavassa, A. and Chornay, N. and Comoretto, G. and Contursi, G. and Cooper, W. J. and Cornez, T. and Cowell, S. and Crifo, F. and Cropper, M. and Crosta, M. and Crowley, C. and Dafonte, C. and Dapergolas, A. and David, M. and David, P. and Laverny, P. de and Luise, F. De and March, R. De and Ridder, J. De and Souza, R. de and Torres, A. de and Peloso, E. F. del and Pozo, E. del and Delbo, M. and Delgado, A. and Delisle, J.-B. and Demouchy, C. and Dharmawardena, T. E. and Matteo, P. Di and Diakite, S. and Diener, C. and Distefano, E. and Dolding, C. and Edvardsson, B. and Enke, H. and Fabre, C. and Fabrizio, M. and Faigler, S. and Fedorets, G. and Fernique, P. and Fienga, A. and Figueras, F. and Fournier, Y. and Fouron, C. and Fragkoudi, F. and Gai, M. and Garcia-Gutierrez, A. and Garcia-Reinaldos, M. and García-Torres, M. and Garofalo, A. and Gavel, A. and Gavras, P. and Gerlach, E. and Geyer, R. and Giacobbe, P. and Gilmore, G. and Girona, S. and Giuffrida, G. and Gomel, R. and Gomez, A. and González-Núñez, J. and González-Santamaría, I. and González-Vidal, J. J. and Granvik, M. and Guillout, P. and Guiraud, J. and Gutiérrez-Sánchez, R. and Guy, L. P. and Hatzidimitriou, D. and Hauser, M. and Haywood, M. and Helmer, A. and Helmi, A. and Sarmiento, M. H. and Hidalgo, S. L. and Hilger, T. and Hładczuk, N. and Hobbs, D. and Holland, G. and Huckle, H. E. and Jardine, K. and Jasniewicz, G. and Piccolo, A. Jean-Antoine and Jiménez-Arranz, Ó and Jorissen, A. and Campillo, J. Juaristi and Julbe, F. and Karbevska, L. and Kervella, P. and Khanna, S. and Kontizas, M. and Kordopatis, G. and Korn, A. J. and Kóspál, Á and Kostrzewa-Rutkowska, Z. and Kruszyńska, K. and Kun, M. and Laizeau, P. and Lambert, S. and Lanza, A. F. and Lasne, Y. and Campion, J.-F. Le and Lebreton, Y. and Lebzelter, T. and Leccia, S. and Leclerc, N. and Lecoeur-Taibi, I. and Liao, S. and Licata, E. L. and Lindstrøm, H. E. P. and Lister, T. A. and Livanou, E. and Lobel, A. and Lorca, A. and Loup, C. and Pardo, P. Madrero and Romeo, A. Magdaleno and Managau, S. and Mann, R. G. and Manteiga, M. and Marchant, J. M. and Marconi, M. and Marcos, J. and Santos, M. M. S. Marcos and Pina, D. Marín and Marinoni, S. and Marocco, F. and Marshall, D. J. and Polo, L. Martin and Martín-Fleitas, J. M. and Marton, G. and Mary, N. and Masip, A. and Massari, D. and Mastrobuono-Battisti, A. and Mazeh, T. and McMillan, P. J. and Messina, S. and Michalik, D. and Millar, N. R. and Mints, A. and Molina, D. and Molinaro, R. and Molnár, L. and Monari, G. and Monguió, M. and Montegriffo, P. and Montero, A. and Mor, R. and Mora, A. and Morbidelli, R. and Morel, T. and Morris, D. and Muraveva, T. and Murphy, C. P. and Musella, I. and Nagy, Z. and Noval, L. and Ocaña, F. and Ogden, A. and Ordenovic, C. and Osinde, J. O. and Pagani, C. and Pagano, I. and Palaversa, L. and Palicio, P. A. and Pallas-Quintela, L. and Panahi, A. and Payne-Wardenaar, S. and Esteller, X. Peñalosa and Penttilä, A. and Pichon, B. and Piersimoni, A. M. and Pineau, F.-X. and Plachy, E. and Plum, G. and Poggio, E. and Prša, A. and Pulone, L. and Racero, E. and Ragaini, S. and Rainer, M. and Raiteri, C. M. and Rambaux, N. and Ramos, P. and Ramos-Lerate, M. and Fiorentin, P. Re and Regibo, S. and Richards, P. J. and Diaz, C. Rios and Ripepi, V. and Riva, A. and Rix, H.-W. and Rixon, G. and Robichon, N. and Robin, A. C. and Robin, C. and Roelens, M. and Rogues, H. R. O. and Rohrbasser, L. and Romero-Gómez, M. and Rowell, N. and Royer, F. and Mieres, D. Ruz and Rybicki, K. A. and Sadowski, G. and Núñez, A. Sáez and Sellés, A. Sagristà and Sahlmann, J. and Salguero, E. and Samaras, N. and Gimenez, V. Sanchez and Sanna, N. and Santoveña, R. and Sarasso, M. and Schultheis, M. and Sciacca, E. and Segol, M. and Segovia, J. C. and Ségransan, D. and Semeux, D. and Shahaf, S. and Siddiqui, H. I. and Siebert, A. and Siltala, L. and Silvelo, A. and Slezak, E. and Slezak, I. and Smart, R. L. and Snaith, O. N. and Solano, E. and Solitro, F. and Souami, D. and Souchay, J. and Spagna, A. and Spina, L. and Spoto, F. and Steele, I. A. and Steidelmüller, H. and Stephenson, C. A. and Süveges, M. and Surdej, J. and Szabados, L. and Szegedi-Elek, E. and Taris, F. and Taylor, M. B. and Teixeira, R. and Tolomei, L. and Tonello, N. and Torra, F. and Torra, J. and Elipe, G. Torralba and Trabucchi, M. and Tsounis, A. T. and Turon, C. and Ulla, A. and Unger, N. and Vaillant, M. V. and Dillen, E. van and Reeven, W. van and Vanel, O. and Vecchiato, A. and Viala, Y. and Vicente, D. and Voutsinas, S. and Weiler, M. and Wevers, T. and Wyrzykowski, Ł and Yoldas, A. and Yvard, P. and Zhao, H. and Zorec, J. and Zucker, S. and Zwitter, T.},
	month = jun,
	year = {2023},
	pages = {A1},
}

@article{rauer_plato_2014,
	title = {The {PLATO} 2.0 mission},
	volume = {38},
	issn = {0922-6435},
	url = {https://ui.adsabs.harvard.edu/abs/2014ExA....38..249R/abstract},
	doi = {10.1007/s10686-014-9383-4},
	language = {en},
	number = {1-2},
	urldate = {2026-03-27},
	journal = {Experimental Astronomy},
	author = {Rauer, H. and Catala, C. and Aerts, C. and Appourchaux, T. and Benz, W. and Brandeker, A. and Christensen-Dalsgaard, J. and Deleuil, M. and Gizon, L. and Goupil, M.-J. and Güdel, M. and Janot-Pacheco, E. and Mas-Hesse, M. and Pagano, I. and Piotto, G. and Pollacco, D. and Santos, Ċ and Smith, A. and Suárez, J.-C. and Szabó, R. and Udry, S. and Adibekyan, V. and Alibert, Y. and Almenara, J.-M. and Amaro-Seoane, P. and Eiff, M. Ammler-von and Asplund, M. and Antonello, E. and Barnes, S. and Baudin, F. and Belkacem, K. and Bergemann, M. and Bihain, G. and Birch, A. C. and Bonfils, X. and Boisse, I. and Bonomo, A. S. and Borsa, F. and Brandão, I. M. and Brocato, E. and Brun, S. and Burleigh, M. and Burston, R. and Cabrera, J. and Cassisi, S. and Chaplin, W. and Charpinet, S. and Chiappini, C. and Church, R. P. and Csizmadia, Sz and Cunha, M. and Damasso, M. and Davies, M. B. and Deeg, H. J. and Díaz, R. F. and Dreizler, S. and Dreyer, C. and Eggenberger, P. and Ehrenreich, D. and Eigmüller, P. and Erikson, A. and Farmer, R. and Feltzing, S. and de Oliveira Fialho, F. and Figueira, P. and Forveille, T. and Fridlund, M. and García, R. A. and Giommi, P. and Giuffrida, G. and Godolt, M. and Gomes da Silva, J. and Granzer, T. and Grenfell, J. L. and Grotsch-Noels, A. and Günther, E. and Haswell, C. A. and Hatzes, A. P. and Hébrard, G. and Hekker, S. and Helled, R. and Heng, K. and Jenkins, J. M. and Johansen, A. and Khodachenko, M. L. and Kislyakova, K. G. and Kley, W. and Kolb, U. and Krivova, N. and Kupka, F. and Lammer, H. and Lanza, A. F. and Lebreton, Y. and Magrin, D. and Marcos-Arenal, P. and Marrese, P. M. and Marques, J. P. and Martins, J. and Mathis, S. and Mathur, S. and Messina, S. and Miglio, A. and Montalban, J. and Montalto, M. and Monteiro, M. J. P. F. G. and Moradi, H. and Moravveji, E. and Mordasini, C. and Morel, T. and Mortier, A. and Nascimbeni, V. and Nelson, R. P. and Nielsen, M. B. and Noack, L. and Norton, A. J. and Ofir, A. and Oshagh, M. and Ouazzani, R.-M. and Pápics, P. and Parro, V. C. and Petit, P. and Plez, B. and Poretti, E. and Quirrenbach, A. and Ragazzoni, R. and Raimondo, G. and Rainer, M. and Reese, D. R. and Redmer, R. and Reffert, S. and Rojas-Ayala, B. and Roxburgh, I. W. and Salmon, S. and Santerne, A. and Schneider, J. and Schou, J. and Schuh, S. and Schunker, H. and Silva-Valio, A. and Silvotti, R. and Skillen, I. and Snellen, I. and Sohl, F. and Sousa, S. G. and Sozzetti, A. and Stello, D. and Strassmeier, K. G. and Švanda, M. and Szabó, Gy M. and Tkachenko, A. and Valencia, D. and Van Grootel, V. and Vauclair, S. D. and Ventura, P. and Wagner, F. W. and Walton, N. A. and Weingrill, J. and Werner, S. C. and Wheatley, P. J. and Zwintz, K.},
	month = nov,
	year = {2014},
	pages = {249--330},
}

@article{gilliland_kepler_2010,
	title = {Kepler {Asteroseismology} {Program}: {Introduction} and {First} {Results}},
	volume = {122},
	issn = {1538-3873},
	shorttitle = {Kepler {Asteroseismology} {Program}},
	url = {https://iopscience.iop.org/article/10.1086/650399},
	doi = {10.1086/650399},
	language = {en},
	number = {888},
	urldate = {2026-03-27},
	journal = {Publications of the Astronomical Society of the Pacific},
	publisher = {IOP Publishing},
	author = {Gilliland, Ronald L. and Brown, Timothy M. and Christensen-Dalsgaard, Jørgen and Kjeldsen, Hans and Aerts, Conny and Appourchaux, Thierry and Basu, Sarbani and Bedding, Timothy R. and Chaplin, William J. and Cunha, Margarida S. and Cat, Peter De and Ridder, Joris De and Guzik, Joyce A. and Handler, Gerald and Kawaler, Steven and Kiss, László and Kolenberg, Katrien and Kurtz, Donald W. and Metcalfe, Travis S. and Monteiro, Mario J. P. F. G. and Szabó, Robert and Arentoft, Torben and Balona, Luis and Debosscher, Jonas and Elsworth, Yvonne P. and Quirion, Pierre-Olivier and Stello, Dennis and Suárez, Juan Carlos and Borucki, William J. and Jenkins, Jon M. and Koch, David and Kondo, Yoji and Latham, David W. and Rowe, Jason F. and Steffen, Jason H.},
	month = jan,
	year = {2010},
	pages = {131},
}

@misc{chen_insights_2026,
	title = {Insights into the {Exoplanet} {Radius} {Valley} from {Host}-{Star} {Ages}, {Activity}, {Chemistry}, and {Birth} {Radius}},
	url = {http://arxiv.org/abs/2603.21581},
	doi = {10.48550/arXiv.2603.21581},
	urldate = {2026-04-09},
	publisher = {arXiv},
	author = {Chen, Xunzhou and Sun, Tiancheng and Yuxi and Lu and Lu, Zixuan and Ye, Lifei},
	month = mar,
	year = {2026},
	note = {arXiv:2603.21581 [astro-ph]},
}
\bibliographystyle{aasjournal}

\end{document}